\documentclass[twocolumns]{aa}  

\usepackage{color}
\usepackage{graphicx}
\usepackage{multirow}
\usepackage{ulem}
\usepackage[bookmarks = true,colorlinks = true,linkcolor = blue,urlcolor = blue,citecolor = blue,pdftex,bookmarks = true,linktocpage = true,hyperindex = true,breaklinks]{hyperref}

\graphicspath{ {./figures/} }
\usepackage{txfonts}
\newcommand{\IRIS}{\textit{IRIS}}

\newcommand{\tauamb}{\tau_{q_{\rm  amb}}} 

\newcommand{\etaamb}{\eta_{_{\rm amb}}}
\newcommand{\etaambb}{\eta^{\ast}_{_{\rm amb}}}
\newcommand{\etahall}{\eta_{_{\rm Hall}}}

\newcommand{\uamb}{\mathbf{u}_{_{\rm amb}}}

\begin{document}

   \title{Nonequilibrium ionization and ambipolar diffusion in solar magnetic flux emergence processes}
   \titlerunning{Nonequilibrium ionization and ambipolar diffusion}

   \author{D. N\'obrega-Siverio\inst{1,2}
          \and
          F. Moreno-Insertis\inst{3,4}
          \and
          J. Mart\'inez-Sykora\inst{5,6,1,2}
          \and
          M. Carlsson\inst{1,2}
          \and
          M. Szydlarski\inst{1,2}
          %\fnmsep\thanks{\email{desiveri@astro.uio.no}}
          }

   \institute{Rosseland Centre for Solar Physics, University of Oslo, PO Box 1029 Blindern, NO-0315 Oslo, Norway\\
   \email{desiveri@astro.uio.no}\\
             \and
             Institute of Theoretical Astrophysics, University of Oslo, PO Box 1029 Blindern, NO-0315 Oslo, Norway\\
             \and
              Instituto de Astrofisica de Canarias, Via Lactea, s/n, E-38205 La Laguna (Tenerife), Spain\\
              \and
              Department of Astrophysics, Universidad de La Laguna, E-38200 La Laguna (Tenerife), Spain\\
              \and 
              Bay Area Environmental Research Institute, NASA Research Park, Moffett Field, CA 94952, USA\\
              \and
              Lockheed Martin Solar and Astrophysics Laboratory, Palo Alto, CA 94304, USA
             }

   \date{Received October 18, 2019; accepted November 27, 2019}

% \abstract{}{}{}{}{} 
% 5 {} token are mandatory
 
  \abstract
{Magnetic flux emergence from the solar interior has been shown to be a key mechanism
  for unleashing a wide variety of phenomena. However, there are still open questions concerning the rise of
  the magnetized plasma through the atmosphere, mainly in the
  chromosphere, where the plasma departs from local thermodynamic equilibrium
  (LTE) and is partially ionized.}
   {We aim to investigate the impact of the nonequilibrium (NEQ)
     ionization and recombination and molecule formation of hydrogen, 
     as well as ambipolar diffusion, on the dynamics and
     thermodynamics of the flux emergence process.}
   {Using the radiation-magnetohydrodynamic Bifrost code, we performed 2.5D
     numerical experiments of magnetic flux emergence from the convection zone up to the corona. The experiments include
     the NEQ ionization and recombination of atomic hydrogen, the
       NEQ formation and dissociation of H$_2$ molecules, and the ambipolar
     diffusion term of the Generalized Ohm's Law.}
   {Our experiments show that the LTE assumption substantially
     underestimates the ionization fraction in most of the emerged region,
     leading to an artificial increase in the ambipolar diffusion and,
     therefore, in the heating and temperatures as compared to
     those found when taking the NEQ effects on the
       hydrogen ion population into account. We see that LTE also overestimates the
     number density of H$_2$ molecules within the emerged region, thus
     mistakenly magnifying the exothermic contribution of the H$_2$ molecule
     formation to the thermal energy during the flux emergence process. We
     find that the ambipolar diffusion does not significantly affect the
     amount of total unsigned emerged magnetic flux, but it is
     important in the shocks that cross the emerged
     region, heating the plasma on characteristic times ranging
     from 0.1 to 100 s. We also briefly discuss the importance of
       including elements heavier than hydrogen in the equation of state
       so as not to overestimate the role of ambipolar diffusion in the atmosphere.}
   {}
   \keywords{Sun: atmosphere --
                Sun: chromosphere --
                Sun: magnetic fields --
                Methods: numerical}
\maketitle
%
%____________________________________________________________________________________________________________________________________
%
% SECTION 1: Introduction
%
%____________________________________________________________________________________________________________________________________
%
\section{Introduction}\label{sec:introduction}
%
%---------------------------------------------------------------------------------------------------
%   Magnetic flux emergence
%--------------------------------------------------------------------------------------------------- 

Magnetic flux emergence is a fundamental process that brings magnetic field
from the solar interior to the atmosphere. It is key not only in
understanding the solar magnetic activity, but also in improving our
predictions of space weather events: many prominent features in the solar
atmosphere are related to this fundamental mechanism. From the theoretical
perspective, magnetic flux emergence has been addressed from different points
of view.  Some authors have focused on the rise of the magnetized plasma
through the convection zone, first by means of idealized magnetohydrodynamics
(MHD for short) experiments of the rise of twisted magnetic tube through
stratified media \citep[see, e.g.,][and references
  therein]{moreno1996,longcope1996,emonet1998,Martinez-Sykora:2015mas}, and
then through radiation-MHD experiments that include a self-consistent
convection zone, \citep[e.g.,][among
  others]{Martinez-Sykora:2008aa,Tortosa2009,MorenoInsertis:2018}.  For
instance, \cite{Cheung:2007aa} showed that granular motions can strongly modify
the rise of the magnetized tubes, deforming, slowing down, and even breaking
them into separate strands. This means that the pattern of arrival of the
magnetized plasma at the surface critically depends on the evolution of the
flux emergence in the solar interior.  Other authors have analyzed the
interaction of the emerging plasma with the preexisting coronal field: from
early numerical simulations by
\cite{forbes1984,Shibata1992a,Yokoyama:1995uq,Yokoyama:1996kx} to the most
recent ones \citep[see, e.g.,
][]{Moreno-Insertis:2013aa,Fang:2014,MacTaggart:2015,Nobrega-Siverio:2016,
Hansteen:2017ib,Ni:2017,Zhao:2018,YangL:2018,Hansteen:2019}.
The interaction between the emerged magnetic plasma and the ambient field can
be manifested in many ways, such as: (a) the impulsive release of mass and energy
that may constitute a significant input to the upper solar atmosphere and to
the solar wind \citep{Raoufai:2016}; (b) the formation of strong shocks and
the generation of turbulence \citep{Priest:2014}; (c) nonthermal processes and
the acceleration of particles \citep{Priest:2002}; (d) and quasi-periodic radio
emission due to tearing instabilities and coalescence of plasmoids in the
current sheet \citep{Karlicky:2010}. The interaction between
the emerged region and the preexisting coronal field can even provide
telltale signatures about the structure of the magnetic fields below the
surface, which is a useful diagnostic tool for the solar dynamo
\citep{Cheung:2014}. All of those implications make magnetic flux emergence a
vibrant and active research area. In spite of the continued theoretical
effort, there are still open questions concerning this fundamental
process. Flux emergence processes occur through various layers in the Sun in
which many different physical mechanisms are involved, and where usually
several assumptions are made to be able to deal with the complexity of those
layers. As \cite{Leenaarts:2018} recently pointed out, flux
emergence numerical experiments including radiation and the interaction
  between ions and neutrals have not been reported so far. In fact,
    basic questions are still unanswered like, for instance, how the energy is
  transported and dissipated in the chromosphere and, more generally, which
  physical ingredients are necessary for a realistic model of emerging flux.

%---------------------------------------------------------------------------------------------------
%   Ambipolar diffusion
%--------------------------------------------------------------------------------------------------- 
A common assumption in numerical experiments of the Sun is to model the plasma using
the MHD approximation in which  the plasma is treated as a 
single fluid with complete coupling between its constituent microscopic species. This 
approximation is able to successfully explain many phenomena in different solar contexts; 
nonetheless, there are regions and phenomena where this assumption is no longer valid  
because, for example, the plasma is partially ionized and there is a decrease in the collisional
coupling \citep[][among others]{Zweibel:2011, Khomenko:2012,Martinez-Sykora2015,Zweibel:2015,Shelyag:2016,Ballester:2018}. 
There is a way to relax the MHD approximations to still treat the plasma as a single fluid 
but including the mentioned effects: the Generalized Ohm's Law 
(\citealp[see the fundamental books by][]{Braginskii:1965,Mitchner:1973,Cowling:1976} 
and its implementation in codes by, e.g., \citealp[][]{Leake:2005rt,osullivan2007,Cheung:2012uq,Martinez-Sykora:2012uq,Gonzalez-Morales:2018}). 
Numerical experiments with this extension report a large impact of the interaction between neutrals 
and ions in the lower solar atmosphere. For instance, this interaction is key to getting type~II spicules and misalignment between the thermal and magnetic structures in the chrosmosphere \citep{Martinez-Sykora:2016,Martinez-Sykora:2017sc,Martinez-Sykora:2017yo}; 
it is able to damp Alfvén waves in the chromosphere \citep{De-Pontieu:2001fj,Leake:2005rt,Soler:2015,Cally:2018,Khomenko:2019}; 
and it also affects the onset of instabilities \citep{Khomenko:2014zr,Ruderman:2018}.
In particular, for magnetic flux emergence processes, the ambipolar diffusion
and the associated Pedersen dissipation have been shown to counteract, to
some extent, the cooling during the expansion of the magnetized plasma in the
atmosphere and to lead to the slippage of the magnetic field with respect to
the bulk plasma velocity, thus increasing the total magnetic flux that
emerges in the solar atmosphere
\citep{Leake:2006kx,Arber:2007yf,Leake:2013dq}. However, those computations
including partial ionization effects were carried out assuming a plasma constituted only by hydrogen and using moreover a simple model based on
the modified Saha equation to calculate
departures from local thermodynamic equilibrium (LTE) instead of a fully
nonequilibrium (NEQ) ionization calculation.

%----------------------------------------------------------------------------------------------------------------- 
%   NEQ
%----------------------------------------------------------------------------------------------------------------- 
Important departures from the ionization equilibrium in the solar atmosphere
have been predicted by theory for several decades now.  The seminal papers by
\cite{Klein:1976,Klein:1978,Kneer:1980,Carlsson:1992,Carlsson:2002wl}, among
others, showed by means of 1D simulations that ionization in shocks occurs on
a faster timescale than recombination behind them. 
%and that in LTE almost all
%the thermal energy goes into ionization 
On the contrary, in LTE, almost all thermal energy is suddenly gone into ionization or taken from recombination, so less temperature increase is reached in the postshocks in comparison with
the NEQ case.  Major improvements in the
computational capabilities have provided
%allowed to overcome some of the limitations of previous numerical studies, thus providing 
a much more complete perspective of the NEQ processes in the chromosphere and transition region.
For example, 
\cite{Leenaarts:2006,Leenaarts:2007,Leenaarts:2011} and
\cite{Golding:2014,Golding:2016} explored, through 2D and 3D numerical
experiments, the large thermodynamical variations in the
chromosphere due to the NEQ ionization and recombination of hydrogen and helium.
From those works, it was concluded that the ionization degree of hydrogen and
helium and, consequently, the electron density cannot be calculated 
in the chromosphere using the LTE approximation.  Other authors have shown that heavy ions
also suffer important departures from LTE, for instance, by means of 1D hydrodynamic
simulations in coronal loops, nanoflares and other impulsive events
\citep[][]{Bradshaw:2003,Bradshaw:2006,Bradshaw:2011,Reep:2017,Reep:2018}; or
in multidimensional radiation-MHD experiments that additionally included
spectral synthesis to explain different observational features in the
transition region
\citep{Olluri:2013fu,olluri:2015,De-Pontieu2015,Martinez-Sykora:2016obs} or
in solar phenomena like surges
\citep{Nobrega-Siverio:2017,Nobrega-Siverio:2018}.  Relaxing the
  LTE condition 
  is therefore important when considering the ionization degree of the most abundant species, especially since the free electron and ion number
  density influence, for example, the radiative losses in the atmosphere.  
\cite{Hansteen:1993} found 
deviations of more than a factor two  in the optically thin losses in a 1D
nanoflare model  when considering NEQ effects. See also the dependence on
the number density of different elements in the chromospheric radiative loss
expressions proposed by \cite{Carlsson:2012uq}.

%----------------------------------------------------------------------------------------------------------------- 
%   LAYOUT
%----------------------------------------------------------------------------------------------------------------- 

The aim of this paper is 
to analyze the role of the
NEQ ionization and recombination and partial
ionization, namely, the effects due to the ion-neutral interactions, during
the emergence of the magnetized plasma in the chromosphere.
The layout of the paper is as follows. Section \ref{sec:model} contains the description 
of the underlying magnetic flux emergence numerical model. Section \ref{sec:emerged_region} shows the main results of the paper, focusing
on the ionization fraction, temperature, and molecular fraction within the emerged region
(Section \ref{sec:emerged_region_properties}), the shocks within the emerged region and the
associated ambipolar diffusion heating (Section \ref{sec:tauqjamb}),
and the amount of magnetic flux that emerges from the solar interior (Section \ref{sec:slippage}).
Finally, Section \ref{sec:conclusions} summarizes and discusses the main conclusions of the present work,
as well as the limitations of the present work.

\begin{figure*}
    \centering
    \includegraphics[width=\textwidth]{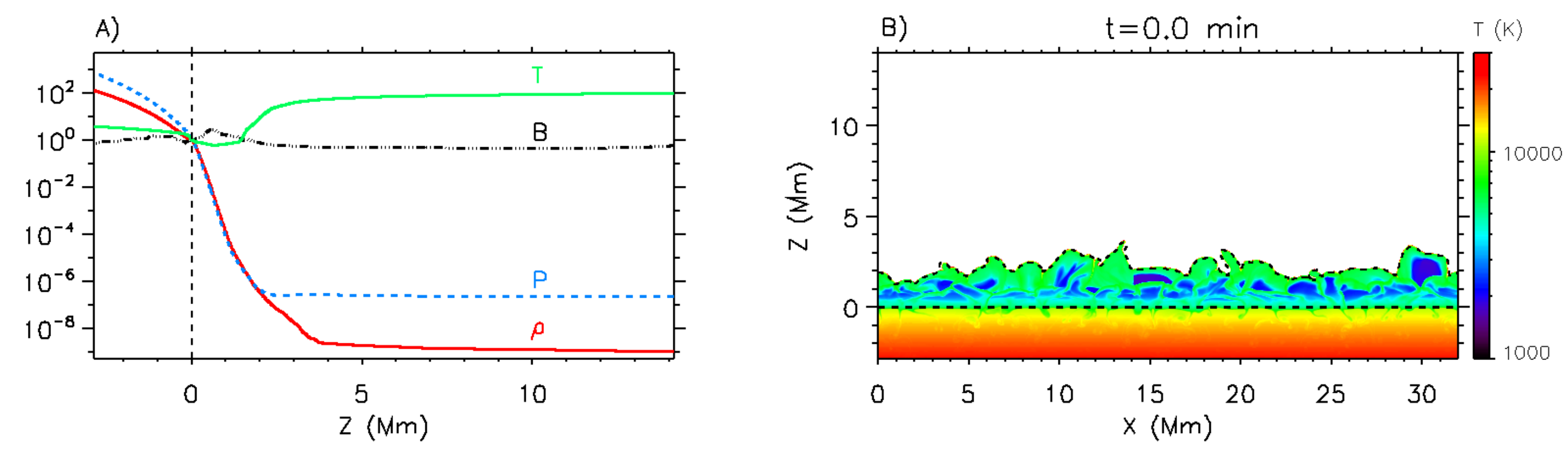}  
    \caption{Properties of the initial snapshot ($t=0$ min). Panel A: Horizontal averages of the initial stratification for temperature $T$ (green), 
    magnetic field strength $B$ (black), 
    pressure $P$ (blue) and density $\rho$ (red) normalized to their photospheric values at $z = 0$~Mm ($T_{\rm ph}= 5803$~K,  $B_{\rm ph}= 0.44$~G,
    $P_{\rm ph} = 1.12\times 10^{5}$~erg cm$^{-3}$, and $\rho_{\rm ph} = 3.20 \times 10^{-7}$~g cm$^{-3}$). 
    The dotted vertical line delineates the solar surface. Panel B: 2D map for the 
    initial background stratification of the temperature only showing values below $T=3\times10^4$ K. The solar surface is roughly at $z=0$~Mm (dashed horizontal black line). The 
    other dashed line is an isocontour that delimits the transition region at $T=10^5$ K.}
    \label{fig:1}  
\end{figure*}

%____________________________________________________________________________________________________________________________________
%
% SECTION 2: Physical and numerical model
%
%____________________________________________________________________________________________________________________________________
\section{Physical and numerical model}\label{sec:model}
For this paper, we have run 2.5D numerical flux emergence 
experiments. The details of the underlying model 
are divided into four sections: the numerical code 
and the specific modules enabled for the present calculations
 (Section \ref{sec:code}); the background 
stratification and the boundary conditions (Section \ref{sec:background}); 
the properties of the twisted magnetic tube 
injected to produce flux emergence (Section \ref{sec:tube});
and the initial stages and branches of the experiments
(Section \ref{sec:branches}).

%____________________________________________________________________________________________________________________________________
\subsection{The numerical code}\label{sec:code}
The numerical experiments were performed using 
the 3D radiation-MHD Bifrost code 
\citep{Gudiksen:2011qy}. This code takes radiative 
transfer with scattering into account 
\citep{Skartlien2000,Hayek:2010ac}, includes the most 
important radiative gains and losses in the chromosphere due to the strong lines from hydrogen, calcium and magnesium 
\citep{Carlsson:2012uq}, 
apart from optically thin losses and thermal conduction along 
the magnetic field in the corona. To prevent the plasma from cooling down 
to low temperatures where the radiation and equation-of-state (EOS) tables 
of the code are not accurate, there is an ad-hoc heating term that forces the 
temperature to stay above $T = 1660$ K. In addition to the above, we use 
the two following modules which are the main ones for this manuscript.

\subsubsection{NEQ ionization recombination and molecule formation of hydrogen}\label{sec:neq}
We have enabled a module for the Bifrost code that computes the 
NEQ ionization and recombination of hydrogen
using a 6-level atom that contains five excitation states for the neutral hydrogen and the ionized level.
This module also calculates the formation and dissociation 
of molecular hydrogen, H$_2$, under NEQ conditions \citep[see][for details about this module]{Leenaarts:2007,Leenaarts:2011}.
    
\subsubsection{Generalized Ohm's Law (GOL)}\label{sec:gol}    
     We have also used a module that extends the classical Ohm's law to include partial ionization effects.
    In particular, we have used a new version of the Generalized Ohm's Law (GOL) module developed by
    \cite{Nobrega-Siverio:2019sts} that improves the capabilities of the original one \citep{Martinez-Sykora:2012uq} by
    implementing the Super-Time-Stepping (STS) method \citep{Alexiades:1996}. This method allows us to relax
    the CFL criterion \citep{Courant:1928uq}, which imposes large
    restrictions on the timestep, to accelerate the explicit calculation of the ambipolar
    diffusion term, which is crucial in magnetic flux emergence
    experiments. In the following, the main equations in this module are briefly summarized.

    In the laboratory reference frame, it can be shown \citep[see, e.g.,][]{Mitchner:1973} that the
    Generalized Ohm's Law (GOL)  is given by 
    \begin{equation}
      {\mathbf E} = - {\mathbf u} \times {\mathbf B} +  \eta {\mathbf J}  -  
              \etaamb \frac{({\mathbf J} \times {\mathbf B})  \times {\mathbf B}}{|\mathbf{B}|^2}  + 
              \etahall \frac{({\mathbf J} \times {\mathbf B})}{|\mathbf{B}|},
      \label{eq:gol}
    \end{equation}
    \noindent
    where ${\mathbf u}$ is the plasma velocity, 
    ${\mathbf E}$ the electric field, ${\mathbf B}$ the magnetic field, and ${\mathbf J}$ the current density 
    all measured in that reference frame. The coefficient 
    $\eta$ is the standard ohmic diffusion given by
    \begin{equation}
       \eta  =  \frac{m_e \nu_{e,ni}}{n_e q_e^2};
    \label{eq:eta_ohm}
    \end{equation}
    \noindent
    the ambipolar diffusion coefficient, $\etaamb$,
    \begin{equation}
      \etaamb = \frac{(\rho_N/\rho)^2|\bf{B}|^2}{\Sigma_n \Sigma_i \rho_n \nu^{\ast}_{ni}}; 
       \label{eq:eta_amb}
    \end{equation}
    \noindent
    and the Hall coefficient, $\etahall$, 
    \begin{equation}
       \etahall = \frac{|\mathbf{B}|}{q_e n_e},
	   \label{eq:eta_hall}
	\end{equation}
	\noindent
    where $q_e$ is the electron charge; $n_e$ the number density of electrons; $m_e$ the electron mass; 
    $\rho_N$ the total neutral mass density obtained from the different neutrals $n$ considered, that is., $\rho_N=\Sigma_n \rho_n$; $\rho$ the total mass density; $\nu_{e,ni}$ the total collision frequency 
    of electrons with neutrals and ions; and $\nu^{\ast}_{ni}$ the reduced
    neutral-ion collision frequency (see, e.g., \citealp{Goodman:2000ys}) given by
    \begin{equation}
	   \nu^{\ast}_{ni} = \frac{m_{ni}}{m_n} n_{_{i}} \sigma_{ni} \left( \frac{8 K_B T}{\pi m_{ni}}\right)^{1/2},
	   \label{eq:nu_ni}
	\end{equation}
	\noindent
    with $n_i$ the ion number density; $K_B$ the
    Boltzmann's constant, $T$ the temperature, and $m_{ni} = m_n
    m_i/(m_n + m_i)$ the reduced mass of the neutral and ion
    species. The temperature-dependent cross section between a given neutral
    and charged particle (ion or electron), $\sigma_{ni}$, is implemented
    following \cite{Vranjes:2013ve} for hydrogen and helium. For the elastic
    cross section for hydrogen protons colliding with H$_2$ molecules we use
    \cite{Krstic:1999}. The rest of the collision cross sections follows the
    same assumptions made by \cite{Vranjes:2008uq} for heavy ions: we take
    the cross section between hydrogen (or helium) and protons multiplied by
    $m_i/m_{_{H}}$ (or $m_i/m_{_{He}}$).
    
In this work, the Hall term was not considered. This term is perpendicular to the
electric current $\mathbf{J}$, so it does not play a direct role in the heating
due to dissipation. Moreover, neglecting this term facilitates the comparison
with previous papers where magnetic flux emergence was studied with ambipolar
diffusion but without the Hall term
\citep{Leake:2006kx,Arber:2007yf,Leake:2013dq}.  With respect to the Ohmic
diffusion, it is negligible in comparison with the numerical diffusion of
the code \citep[see, e.g.,][for a comparison of the different GOL
  terms]{Martinez-Sykora:2012uq, Martinez-Sykora:2017yo}. For this reason,
Ohm's term is not included in Bifrost: neither in the classical Ohm's law nor
in the GOL extension. Instead, a hyper-diffusion term, $\eta_{\rm hyp} {\mathbf
  J}$, is implemented, where $\eta_{\rm hyp}$ is the hyper-diffusion
coefficient \citep[for details, see][]{Gudiksen:2011qy}.

%____________________________________________________________________________________________________________________________________
\subsection{The background stratification and boundary conditions}\label{sec:background}
We started from a statistically stationary two-dimensional snapshot
that
encompasses from the uppermost layers of the solar convection zone to the
corona. The relaxation of this snapshot was carried out including all the
physics mentioned above, with the exception of the NEQ module. The latter is
enabled coinciding with the injection of the magnetic twisted tube
through the lower boundary (see Section
\ref{sec:branches} for details).

The physical domain of the numerical box is $0  \leq x \leq 32 $~Mm
and $-2.87  \leq z \leq 14.2 $~Mm, where $z = 0$~Mm corresponds to 
the horizontal layer where $<\tau_{500}>\approx 1$, with $\tau_{500}$ being the optical depth at 500~nm. The domain is solved with $2048\times1000$ grid cells 
using a uniform numerical grid in both the horizontal and vertical directions
with $\Delta x \approx 16 $~km and $\Delta z \approx 17 $~km, respectively. Concerning the boundary
conditions, they are periodic in the horizontal direction. For the vertical direction,
the bottom boundary 
is open and prescribes constant entropy for the incoming plasma; and the top one uses 
characteristic boundary conditions \citep[see][]{Gudiksen:2011qy}. Furthermore, in the corona we have added 
a {\it hot-plate} that forces a fixed temperature in the top cells. 
In this case, we fix the temperature to stay around  $6\times10^5$~K.  This value could seem low for the corona, 
but since we are interested in the details within the emerged region, the results are not affected by this fact.  

Panel A in Figure \ref{fig:1} shows the horizontal averages of the stratification of the statistically stationary initial snapshot
for temperature $T$ (green), magnetic field $B$ (black), gas pressure $P$ (blue), and density $\rho$ (red). 
The stratification curves in the figure are normalized to the photospheric values at $z = 0$~Mm, namely, 
$T_{\rm ph}= 5803$~K, $B_{\rm ph}= 0.44$~G, $P_{\rm ph} = 1.12\times 10^{5}$~erg~cm$^{-3}$, and $\rho_{\rm ph} = 3.20 \times 10^{-7}$~g~cm$^{-3}$. 
Panel B of that figure contains a 2D temperature map for the initial snapshot ($t=0$~min). 
In this initial snapshot we have chosen the magnetic field to be very weak: (1) to allow an easier analysis, since the only magnetized plasma in the atmosphere is 
the one that has emerged; and (2) to prevent any important magnetic reconnection episode between the emerged plasma and the preexisting ambient field 
in the atmosphere, so no hot ejections, surges or other eruptive and explosive phenomena can perturb the emerged region.

%____________________________________________________________________________________________________________________________________
\subsection{The twisted magnetic tube}\label{sec:tube}
In order to produce flux emergence,  
we injected in the initial snapshot ($t=0$~min) a twisted magnetic tube.
The axis of the tube points in the $y$-direction, and the 
longitudinal and transverse components of the magnetic field 
have the following canonical form (see, e.g., \citealp{fan2001}):
\begin{equation}
    B_y = B_0 \, \exp \left( - \frac{r^2}{R_0^2}\right)
    \label{eq:tube1}
\end{equation}
\begin{equation}
    B_{\theta} =  q \, r \, B_y,
    \label{eq:tube2}
\end{equation}
\noindent where $B_0$ is the magnetic field in the axis of the tube, 
$q$ a constant twist parameter, $r$ and $\theta$ are, respectively, the radial and 
azimuthal coordinates with respect to the tube axis, and $R_0$ is the 
tube radius.

The tube is injected
through the bottom boundary following the method described by 
\cite{Martinez-Sykora:2008aa}. 
This method prescribes the magnetic field at the boundaries, updating the height of the tube every timestep according to the average vertical speed of the inflow plasma, $u_z$, where the tube is located. The electric field of the tube is then computed following Ohm's law (e.g., for the x-component $E_x = B_y \ u_z$) preserving the solenoidality condition $\nabla \cdot {\bf B}=0$.

We set up two flux emergence experiments with
the same parameters for the magnetic tube, namely, $B_0=20$~kG,
$q=2.4$~Mm$^{-1}$, and $R_0=0.1$~Mm. Those parameters were selected to get 
an initial axial magnetic flux of $\Phi_0 = 6.3 \times 10^{18}$~Mx,
which is in the lower range of an ephemeral active region \citep{Zwaan:1987yf},
and leads to a coherent emergence pattern at the surface.
The initial height of the tube is set at $z_0=-3.1$ Mm for both experiments. The only difference 
between the two experiments is the horizontal location $x_0$ where the tube is injected (see also first column in Table \ref{table:1}).
This is because the pattern of arrival of the magnetized plasma at the surface  critically depends on the interaction of the tube with the cells in the interior and, 
hence, on $x_0$. For Experiment 1 we used $x_0=12.5$~Mm. For Experiment 2 we used $x_0=13.0$~Mm.

%____________________________________________________________________________________________________________________________________
\subsection{Initial stages and branches of the experiments}\label{sec:branches}
The experiments begin with the injection of the twisted magnetic tube 
through the bottom boundary of the numerical box ($t=0$~min). In that instant, we
switched on the NEQ module. Since the tube is injected deep enough in the convection zone, all the 
transients in the atmosphere related to enabling the NEQ module would have disappeared 
before the tube reaches the surface: the timescales of those transients are small \citep[about a hundred seconds][]{Leenaarts:2007} compared with the time that the twisted tube takes to rise through the convection zone (tens of minutes).
In addition, the NEQ and the ambipolar 
diffusion effects are negligible in the convection zone, so the rise and expansion 
of the magnetic tube in this region are similar to our previous
experiments where we did not include any of those
physical mechanisms \citep{Nobrega-Siverio:2016,Nobrega-Siverio:2018}, namely, 
the tube rises with velocities of a few km~s$^{-1}$; 
the convection flows deform and break the twisted magnetic tube 
into smaller fragments; and, finally, part of the twisted tube 
reaches the surface. In the experiments of the present paper, the rise from 
the injection point up to the surface takes roughly 50 minutes. There, 
the magnetized plasma starts to pile up producing anomalous magnetized granules.
To understand the NEQ and ambipolar diffusion effects in the subsequent
evolution, from $t=60$ min ondward, for both experiments we create the following four branches:
\begin{itemize}
    \item NEQ+AD: here we continue the experiment without changes, in other words, using the NEQ and ambipolar diffusion modules.
    \item NEQ: in this branch we turn off the ambipolar diffusion.
    \item LTE+AD: here we turn off the NEQ module but keep the ambipolar diffusion.
    \item LTE: in this branch we use neither the NEQ nor the ambipolar diffusion modules, so the ionization is computed under the LTE assumption and we use the classical Ohm's law.
\end{itemize}
This way we can separate and understand the role of the different physical mechanisms in the flux emergence process. The instant to create those branches has been carefully chosen as close as possible to the beginning of the phase of emergence through the chromosphere, in order to have similar structures in the preexisting atmosphere and thus be able to perform one-to-one comparisons between branches. Also we have tried to start the branching not too late in time for the NEQ and ambipolar diffusion effects to be negligible in the emerging plasma, and so avoid affecting the subsequent results with our choice (See Appendix \ref{sec:appendix} for further details). Disabling those modules introduces some transients in the chromosphere; nonetheless they vanish faster
than the timescale for the emergence process within the chromosphere.
The evolution of the experiments is summarized in Table \ref{table:1}.

\begin{table*}
\caption{Scheme of the time evolution for the two experiments and the physical mechanisms included at the different stages.}             
\label{table:1}      
\centering          
\begin{tabular}{|| c | c | c | c ||}     % 4 columns 
\hline\hline
\rule{0pt}{2.5ex} 
 Injection of  & Emergence through & 
 Piling up and & Emergence in the\\
  the magnetic tube & the convection zone & 
 buoyancy instability &  higher atmosphere \\
 ($t=0$ min) &  ($0 < t \leq 50$  min) & ($50 < t < 60$ min) &  ($t\geq60$ min) \\
\hline\hline
\rule{0pt}{2ex}  
    \multirow{3}{*}{Experiment 1} & \multirow{4}{*}{NEQ+AD}  & \multirow{4}{*}{NEQ+AD} & LTE  \\ 
     &   &  & LTE+AD \\ 
     ($x_0=12.5$ Mm)  &   &  & NEQ \\ 
      &   &  & NEQ+AD \\ 
   \hline\hline
    \multirow{3}{*}{Experiment 2} & \multirow{4}{*}{NEQ+AD}  & \multirow{4}{*}{NEQ+AD} & LTE  \\ 
     &   &  & LTE+AD  \\ 
     ($x_0=13.0$ Mm) &   &  & NEQ   \\ 
       &   &  & NEQ+AD   \\ 
\hline                  
\end{tabular}
\end{table*}

%____________________________________________________________________________________________________________________________________
%
% SECTION 4: Results
%
%____________________________________________________________________________________________________________________________________
\section{Results}\label{sec:emerged_region}

\begin{figure*}[!ht]
    \centering
    \includegraphics[width=\textwidth]{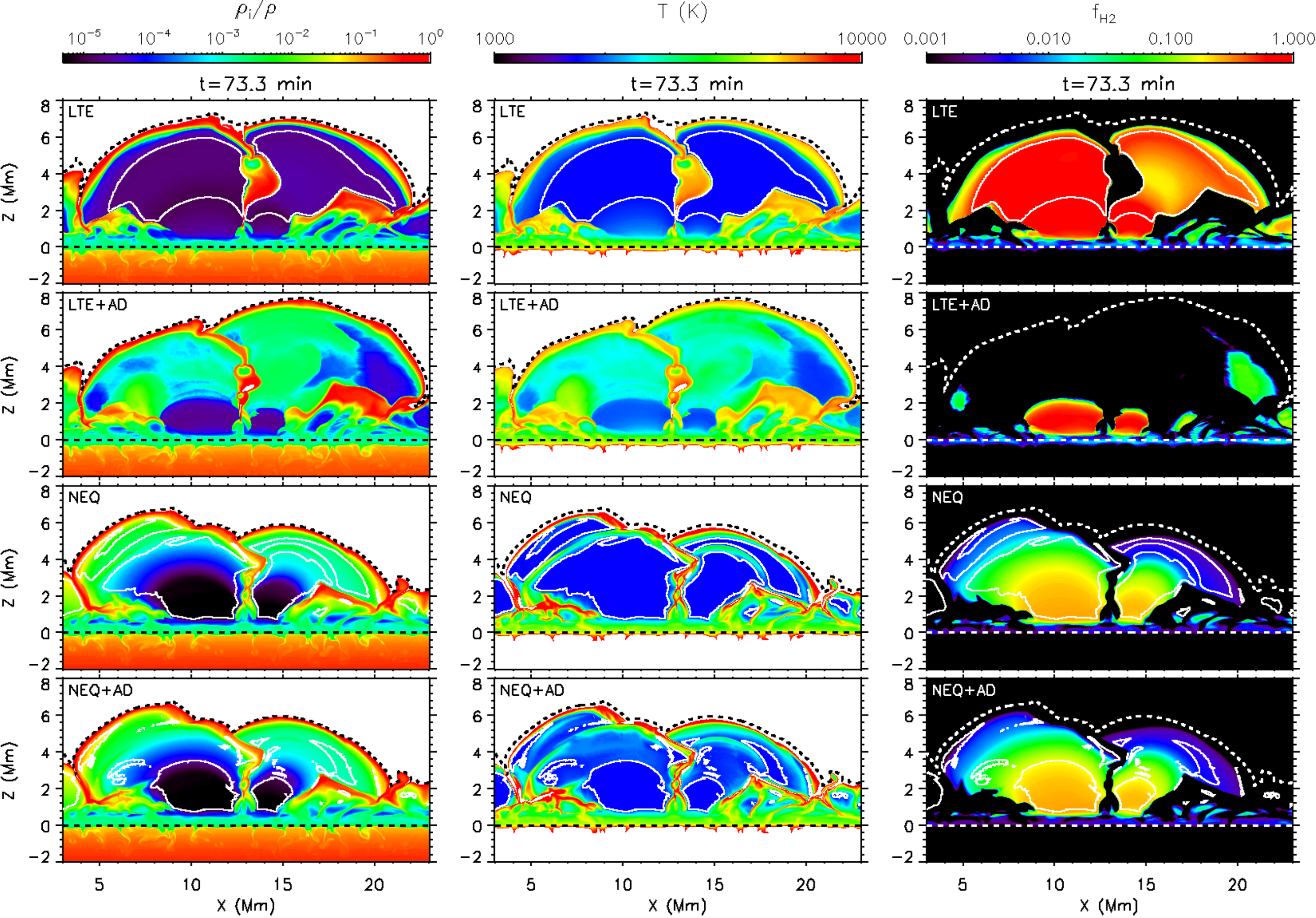}  
    \caption{Emerged region for Experiment 1.  The rows contain the four
      branches of this experiment, namely, in descending order, LTE, LTE+AD,
      NEQ, NEQ+AD. The columns show, from left to right, the ionization
      fraction, $\rho_i/\rho$; the temperature, $T$; and the fraction of
      molecular hydrogen, f$_{\rm H_2}$.  Dashed lines
    indicate the location of  the solar surface (line at
    $z=0$ Mm) and of the transition region (isocontour at $T=10^5$ K).  Solid white isocontours delimit the
      temperature threshold limit at $T=1660$ K.  An animation of this figure
      is available online.}
    \label{fig:2}  
\end{figure*}

When the magnetized plasma of the twisted tube
  reaches the surface, the magnetic field piles up there and the magnetic
  pressure increases. Its later evolution occurs through the development of
  the buoyancy instability \citep{Newcomb:1961aa}, which allows the magnetized plasma to
rise well above the photospheric heights. The magnetized plasma thus
expands into the atmosphere giving rise to dome-like 
structures akin in shape to the ones in previous experiments
\citep[e.g.,][]{archontis2004,moreno-insertis_flux_2006,Nobrega-Siverio:2018}.
In our case, Experiment 1 leads to two emerged domes that interact with each
other, similarly to, for example, \cite{Hansteen:2019}; while Experiment 2 gives rise to a single 
dome as in the paper by \cite{Nobrega-Siverio:2016}. In the following, we analyze the emerged 
region for the four branches (LTE, LTE+AD, NEQ, NEQ+AD) of the two experiments.

\begin{figure*}[!ht]
    \centering
    \includegraphics[width=\textwidth]{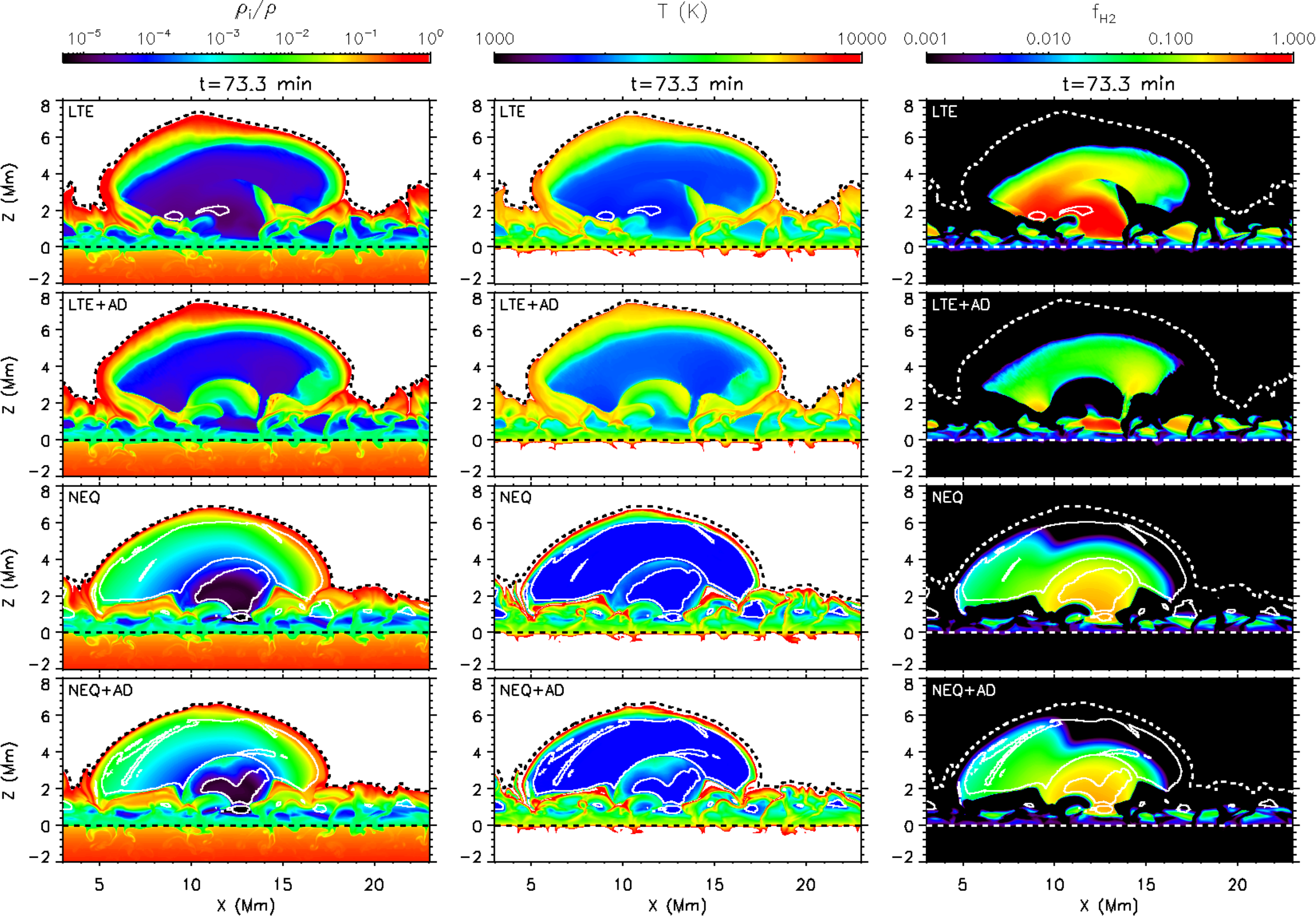}  
    \caption{Emerged region for Experiment 2. 
    The rows contain the four branches of this experiment, namely, in descending order, LTE, LTE+AD, NEQ, NEQ+AD. The
    columns show, from left to right, the ionization fraction, $\rho_i/\rho$; the temperature, $T$; and the fraction of molecular hydrogen, f$_{\rm H_2}$.
    Dashed lines
    indicate the location of  the solar surface (line at
    $z=0$ Mm) and of the transition region (isocontour at $T=10^5$ K).
    Solid white isocontours delimit the temperature threshold limit at $T=1660$ K.
    An animation of this figure is available online.}
    \label{fig:3}  
\end{figure*}

%____________________________________________________________________________________________________________________________________
\subsection{Ionization fraction, temperature, and molecular fraction in the emerged region}\label{sec:emerged_region_properties}
In order to analyze the role of both the 
NEQ populations of atomic and molecular hydrogen
and the ambipolar diffusion, we first focus on a) the ionization fraction,
$\rho_i/\rho$, which is a good tracer of the relevance
of the nonequilibrium ionization and recombination as well as of the ambipolar
diffusion term since its coefficient 
depends on the number density of neutrals and ions (see Equation \ref{eq:eta_amb});
b) the temperature, $T$; and c) the fraction of molecular hydrogen, f$_{\rm H_2}$,  given by
\begin{equation}
    f_{\rm H_2} = \frac{2 n_{\rm H_2}}{\sum_{i=1}^{6}{n_i} + 2 n_{\rm H_2}},
\end{equation}
where we consider that our hydrogen atom model consists of 6 levels (see Section \ref{sec:neq}). 
Figures \ref{fig:2} and \ref{fig:3} show the emerged region for Experiment 1 and 2, respectively.
The four rows of the images are for the different branches (LTE, LTE+AD, NEQ, and NEQ+AD), 
while the three columns contain, from left to right, $\rho_i/\rho$, $T$, and f$_{\rm H_2}$.

    %-------------------------------------------------------------------------------------------------------------------------
    % LTE
    %-------------------------------------------------------------------------------------------------------------------------

    The LTE case (first row of Figures \ref{fig:2} and \ref{fig:3}, and their respectively online animations) shows an extremely low degree of 
    ionization ($\rho_i/\rho < 10^{-4}$) in most of the interior of the domes, except for 
    a vertical region in Experiment 1, where the two emerged domes are colliding and a current 
    sheet is being created between them; and in Experiment 2 around $x=15$
    Mm, in the shock front explained in Section \ref{sec:tauqjamb}.
    % LTE temperature
    Concerning the temperature, the emerged domes suffer strong cooling during their expansion
    with the result that, in Experiment 1, most of the dome is at
    the threshold temperature value where ad-hoc heating is activated (see
    white solid isocontours at $T=1660$~K). Experiment 2 also shows temperatures
    at that threshold but to a lesser extent.
    % LTE H2 molecule 
    With respect to the H$_2$ molecules, we find that they
    constitute around $70$\% of the total hydrogen content by mass within
    the emerged region.  This means that for an LTE case, not even the
    high exothermic contribution of the H$_2$ molecule formation (4.48 eV
    per molecule) is able to counteract the cooling during the flux
    emergence through the atmosphere.
    
    %-------------------------------------------------------------------------------------------------------------------------
    % LTE+AD
    %-------------------------------------------------------------------------------------------------------------------------
    % LTE+AD ionization fraction
    When including ambipolar diffusion in the LTE case (second row of Figures \ref{fig:2} and \ref{fig:3}), 
    we notice several differences. The ionization fraction within the domes
    is now much larger than in the previous case: 
    in Experiment 1, this is evident in almost the whole emerged region; while in Experiment 2, 
    it is visible in a bubble-like structure in the inner core. The reason for this increase in both experiments is the role
    of the ambipolar diffusion related to the shocks that pass through the domes and that we describe in Section \ref{sec:tauqjamb}.
    % LTE+AD temperature
    The temperature in this branch is also greater; in fact, we get temperatures at least $100$~K above the 
    threshold limit during the whole process of emergence (as we can note by the lack of white solid contours at $T=1660$~K 
    in comparison with other branches). In this case, the ad-hoc heating is not necessary at any stage of the flux emergence process.
    % LTE+AD H2 molecule
    With respect to f$_{\rm H_2}$, as a consequence of the larger
    temperatures in comparison with the LTE branch, the formation of
    molecules is substantially reduced in most of the emerged region. In fact, the molecules are limited to
    small regions in the core of the domes; the maximum value of $f_{\rm H_2}$ is around 60\%.
    
    %-------------------------------------------------------------------------------------------------------------------------
    % NEQ
    %-------------------------------------------------------------------------------------------------------------------------
    % NEQ ionization fraction
    The NEQ branch (third row of Figures \ref{fig:2} and \ref{fig:3}) shows a larger $\rho_i/\rho$ 
    in most of the emerged region than the LTE branch, meaning that the assumption of LTE leads to wrong values of 
    the real ionization fraction. This result also impacts on the calculation of the ambipolar diffusion coefficient
    since $\etaamb \propto n_i^{-1}$: in the upper part of the domes, the 
    assumption of LTE overestimates $n_i$ and hence the role of ambipolar diffusion.
    % NEQ temperature 
    Regarding the temperature, in the NEQ branch the threshold limit
    $T=1660$~K is reached to an even larger extent than in the LTE branch,
    especially in Experiment 2. In both experiments we see prominent arch-like
    structures with $T\approx3\times10^3$~K embedded in the cool $T=1660$~K
    domain. They correspond to shock fronts; their
    importance is discussed in the next section.
    % NEQ H2 molecule
    With respect to the molecules, in NEQ there are more ions
    so the formation rate of H$_2$ is less important than in LTE. 
    The maximum molecular fraction f$_{\rm H_2}$ is around 27\% and only
    located in the inner part of the domes. This implies that 
    the LTE assumption vastly overestimates the number density of molecules
    and, therefore, their corresponding thermostatic action
    in magnetic flux emergence episodes.
    
    %-------------------------------------------------------------------------------------------------------------------------
    % NEQ + AD
    %-------------------------------------------------------------------------------------------------------------------------
    % NEQ+AD ionization fraction and H2 molecule
    For the NEQ+AD case (fourth row of Figures \ref{fig:2} and \ref{fig:3}), 
    the ionization and molecular fraction are practically identical to the
    NEQ branch without ambipolar diffusion just discussed. 
    % NEQ+AD temperature
    The main difference between NEQ+AD and NEQ can be found in the areas surrounding the arch-like structures 
    associated with shocks mentioned before. The inclusion of ambipolar diffusion seems to offset some of the 
    strong cooling during the expansion of the emerged region; however it is not enough to avoid reaching the temperature 
    threshold at $T=1660$~K, and, therefore, those domes are affected by the ad-hoc heating.

\begin{figure*}[ht]
    \centering
    \includegraphics[width=0.96\textwidth]{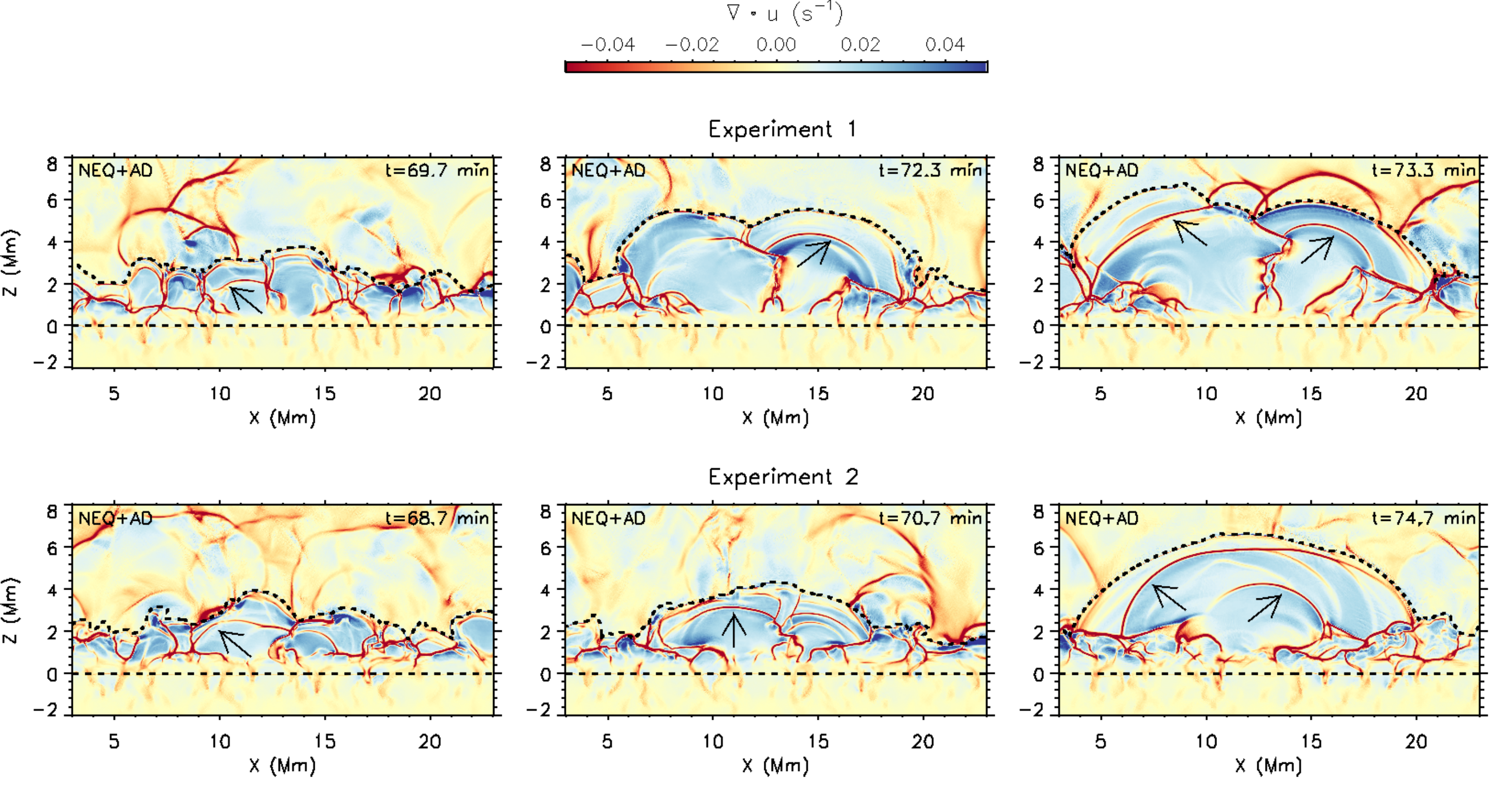}  
    \caption{Divergence of the plasma velocity field, $\nabla \cdot \mathbf{u}$, 
    for experiments 1 (top) and 2 (bottom) at different stages of the magnetic flux emergence process.
    Arrows are superimposed pointing out the location of some of the shocks within the emerged regions.
    Dashed lines
    indicate the location of  the solar surface (line at
    $z=0$ Mm) and of the transition region (isocontour at $T=10^5$ K).
    An animation of this figure is available online.}
    \label{fig:4}  
\end{figure*}

\begin{figure*}[!ht]
    \centering
    \includegraphics[width=0.98\textwidth]{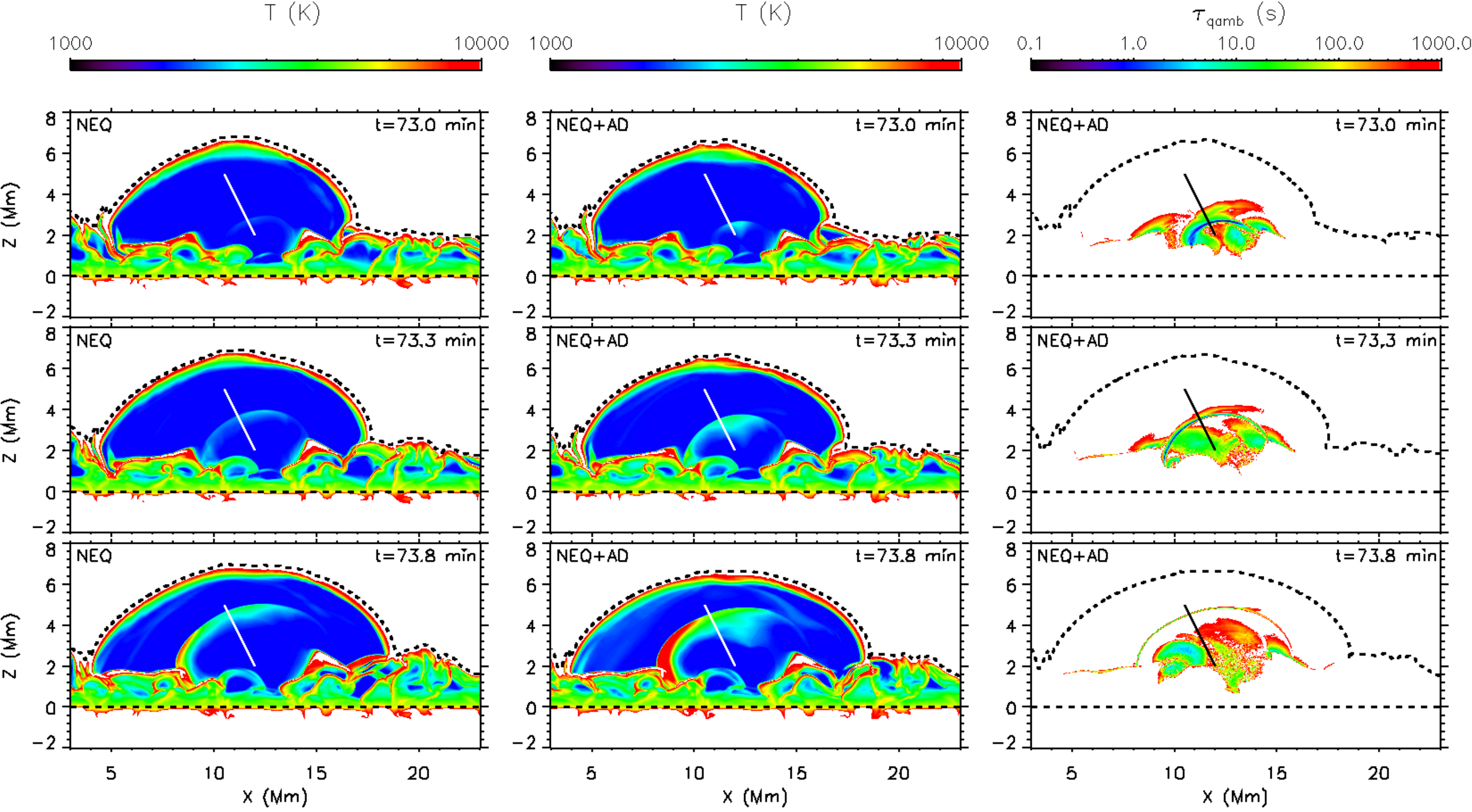}  
    \caption{Shock evolution within the emerged region for the Experiment 2 at different instants (rows). 
    The left column shows the temperature, $T$, for the NEQ branch;
    the middle column contains $T$ for the NEQ+AD case; and the right column illustrates the characteristic time of the heating due to
    ambipolar diffusion, $\tau_{q_{\rm amb}}$ (see Equation \ref{eq:tau_qamb}). In addition, a solid line nearly perpendicular to the shock front 
    has been superimposed in all the panels to study the variation of different quantities due to the shock passage (see Figure \ref{fig:6}).
    In the image, dashed lines
    indicate the location of  the solar surface (line at
    $z=0$ Mm) and of the transition region (isocontour at $T=10^5$ K).
    An animation of this figure is available online.}
    \label{fig:5}  
\end{figure*}
\begin{figure*}[!ht]
    \centering
    \includegraphics[width=1.0\textwidth]{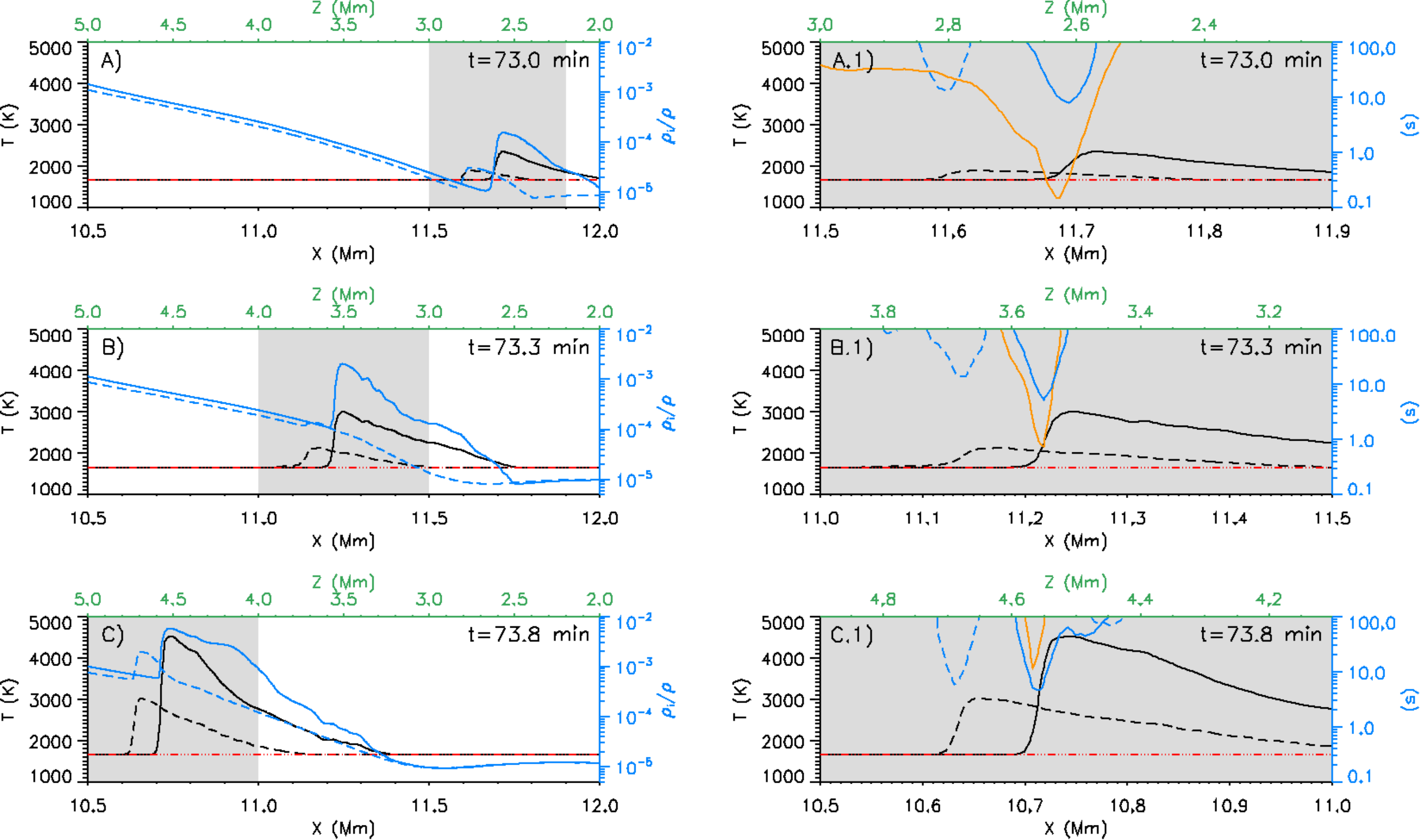}
    \caption{Evolution of different quantities along the cut plotted in Figure \ref{fig:5} for the 
     same instants reproduced in that image. The $x$ and $z$ coordinates along that line are given as abscissas 
     in the bottom and top axes, respectively.
    The left panels contain the evolution of the temperature $T$ (black lines, left axis) and ionization fraction
    $\rho_i/\rho$ (blue lines, right axis) for the NEQ case (long dashed lines) and NEQ+AD (solid ones). 
    The right panels contain a zoom of
    the gray region of the left panels. The curves are for the
    temperature (black), and the characteristic times $\tau_{comp}$ (blue)
    and $\tauamb$ (orange), keeping the long dashed lines for the purely NEQ case and
    solid lines for the NEQ + AD one.  The tick-marks for the temperature are given on the
    left axis, and those for the times on the right axis. For all panels, the threshold temperature, 
    $T=1660$ K, is shown with a horizontal dash-dotted line in red.}
    \label{fig:6}  
\end{figure*}

%____________________________________________________________________________________________________________________________________
\subsection{Shocks within the emerged region and the role of the ambipolar diffusion heating}\label{sec:tauqjamb}

To locate the shock fronts propagating in the domain, in particular those 
already mentioned in the previous paragraphs, we use the divergence of the 
plasma velocity field $\nabla \cdot \mathbf{u}$ as a marker.
In Figure \ref{fig:4} (and associated movie), we show this quantity for the NEQ+AD branch
of both experiments at different stages of the evolution. In the
  image, arrows have been added that point to specific cases of interest in
  the emerged region. In all the branches of the two experiments,
during the rise of the magnetized plasma through the solar corona, 
several compression waves go through the emerged domes and steepen into shocks. 
Their origin is diverse:  the turbulent motion of the convection zone, 
secondary magnetic flux emergence episodes, even the non-stationary reconnection between domes (Experiment 1). 
We have found that the strongest shocks leave an important imprint
in the emerged dome, especially in the experiment branches that take the ambipolar diffusion into account.
To show this in detail we use Experiment 2, in which the
  structure of the shock is simpler. Figure \ref{fig:5} contains the emerged region for that 
experiment at different instants (rows). 
The left column of the image shows the temperature $T$ for the NEQ branch;
the middle column contains $T$ for the NEQ+AD case; and the right column shows, also for NEQ+AD, 
the characteristic time of 
the ambipolar diffusion heating $\tau_{q_{\rm amb}}$, which is defined by
\begin{equation}
    \tau_{q_{\rm  amb}} = \frac{e}{q_{\rm  amb}} = \frac{e}{\etaamb J^2_{\perp}},
    \label{eq:tau_qamb}
\end{equation}
where $e$ is the internal energy per unit volume, $q_{\rm amb}$ is the ambipolar diffusion heating, and $J_{\perp}$ is the 
current perpendicular to the magnetic field. 

Taking a look at the left and middle columns of Figure \ref{fig:5}, 
we can appreciate that there is a prominent front that traverses the dome
and, 
in spite of the energy expended in dissociation and ionization processes in
the shock, results in a marked temperature jump between the pre-shock and
post-shock regions (see also the
  associated animation). The temperature increase
in the NEQ+AD branch is greater, and the width of the 
  relaxation region behind the shock wider,
than in the purely NEQ branch. 
In fact, in the right column of Figure \ref{fig:5}, we can see that the characteristic times of the ambipolar diffusion
heating $\tau_{q_{\rm  amb}}$ associated with the shock are small compared to the evolutionary timescales of the emergence process, 
implying that the heating is very efficient.  In order to better illustrate
this, a slanted solid line, which is nearly perpendicular to the shock front 
at the different instants, has been superimposed in all the panels of Figure
\ref{fig:5}. Thus, we can analyze the profiles of different quantities along that cut, so as to quantify 
the significance of the shock. The corresponding profiles are shown in Figure \ref{fig:6}.
The left column, panels A, B and C, of the image contains the evolution of the temperature $T$ (black lines, left axis) and the ionization fraction
$\rho_i/\rho$ (blue lines, right axis) for the NEQ case (long dashed lines) and NEQ+AD (solid ones). Additionally,
the threshold temperature where the ad-hoc heating is activated ($T=1660$ K) is shown with a horizontal dash-dotted line in red.
In those panels, it is clear that 
the jumps are larger when including ambipolar diffusion;
they can be 
up to a factor two greater for the NEQ+AD case as compared to that with only NEQ.
For the ionization fraction, the jump
is even greater, reaching roughly a factor of one order of magnitude. 
The shock in either branch moves with an average speed of $\sim18$~km
s$^{-1}$. The panels in the right column of Figure \ref{fig:6} (A.1, B.1 and C.1) 
contain a zoom of the gray region of the left panels, showing again the evolution of
the temperature in the shock (for context purposes), but now in the right
axis we plot the characteristic time of the ambipolar diffusion heating,
$\tau_{q_{\rm amb}}$ (orange), together with the characteristic
compression time of the shock, $\tau_{\rm comp}$ (blue lines), given by
\begin{equation}
    \tau_{\rm comp} =  \frac{1}{ | \nabla \cdot \mathbf{u} | }, 
    \label{eq:tau_comp}
\end{equation}
\noindent (in the
plot, only the values in compression regions, $\nabla\cdot\mathbf{u} <0$,  are
shown). 
In all cases, the compression time $\tau_{comp}$ is within a factor
  two of $10$~s, on the order of the time of passage of the plasma through
  the shock. On the other hand, in the first panel of the right column ($t=73.0$ min),
  the ambipolar diffusion heating time, $\tauamb$, is as small as $\sim0.1$
  s, almost two orders of magnitude smaller than $\tau_{comp}$:
    this indicates that the ambipolar diffusion clearly plays a role in
    increasing the internal energy of the plasma as it crosses the shock, even if part of that
    increase is consumed in hydrogen ionization or molecule dissociation
    rather than in raising the temperature.  Eighteen seconds later (at
  $t=73.3$ min), $\tauamb$ is still below $1$~s, that is,
  about one order of magnitude shorter than $\tau_{comp}$. It is only in the
  third panel of the right column (at $t=73.8$ min) that $\tauamb$ is
  near to (even though still smaller than) $\tau_{comp}$.  The short characteristic values for
  $\tauamb$ obtained in the NEQ+AD case explain the significant departures
  seen in the thermal evolution across the shock in that branch with respect to the NEQ
  branch. Considering the passage of the shock
  through the whole dome, which lasts for some $\sim 100$~s, we expect the
  effects of the ambipolar diffusion to be noticeable in changing the
  temperature, and the molecular and atomic H fractions, of the plasma, as we
  have seen to be the case in the previous section and figures.

%____________________________________________________________________________________________________________________________________
\subsection{Magnetic field slippage}\label{sec:slippage}
\begin{figure}[!ht]
    \centering
    \includegraphics[width=0.45\textwidth]{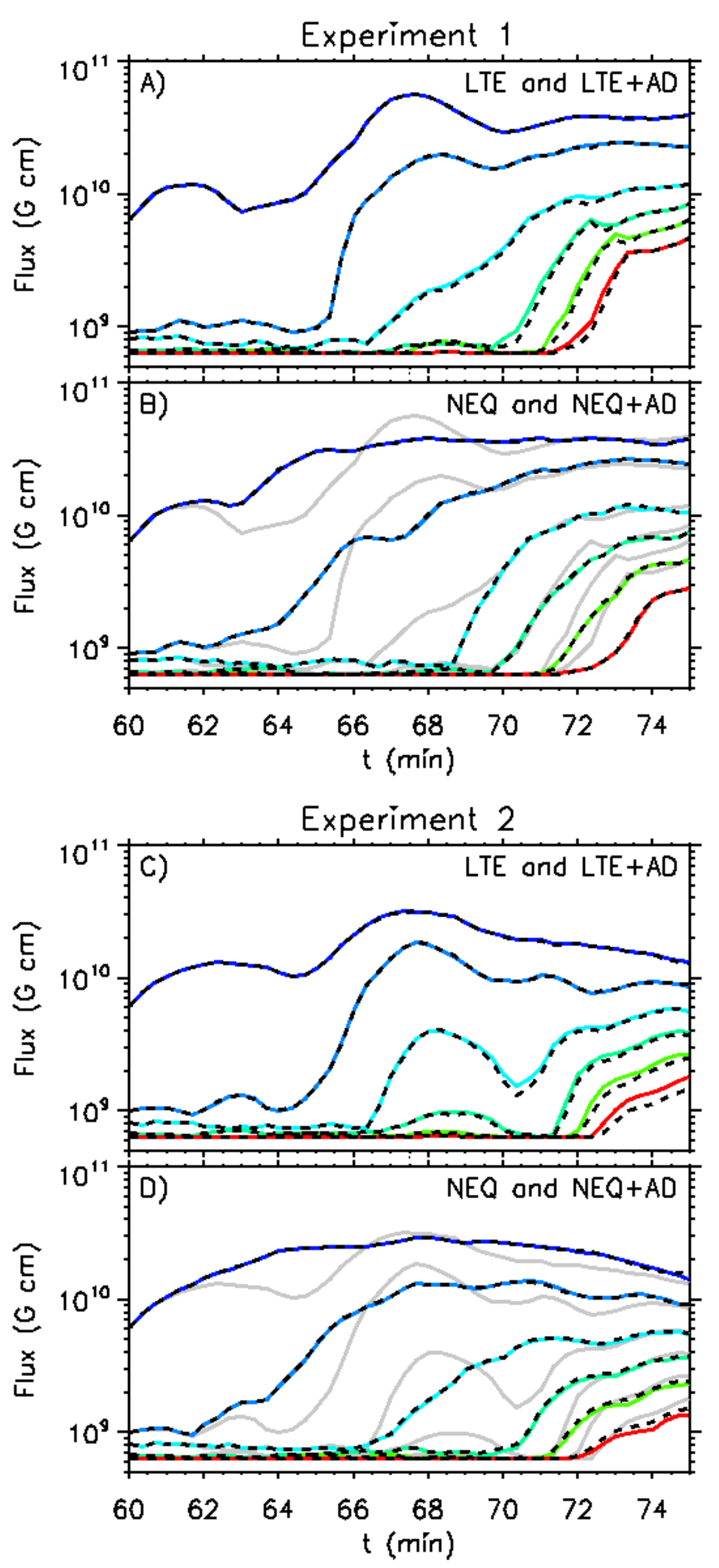}
    \caption{Unsigned vertical flux $\Phi$ as a function of time during the flux emergence process 
    for Experiment 1 (panels A and B) and Experiment 2 (panels C and D) for different heights, namely, $z=[0.5,1.0,2.0,3.0,4.0,5.0]$ Mm. 
    The LTE panel for each experiment contains $\Phi^{\rm AD}_{\rm LTE}$ (colored lines) and $\Phi_{\rm LTE}$ (superimposed dashed lines). 
    The NEQ panels show $\Phi^{\rm AD}_{\rm NEQ}$ (colored lines), $\Phi_{\rm NEQ}$ 
    superimposed dashed lines) and also $\Phi^{\rm AD}_{\rm LTE}$ (gray lines) to compare with the NEQ cases.}
    \label{fig:7}  
\end{figure}

Hitherto, we have studied the dissipation due to ambipolar diffusion and how it affects the temperature. In this section, inspired by previous numerical papers about flux emergence including ion-neutral interaction effects \citep{Leake:2006kx,Arber:2007yf,Leake:2013dq}, we study a second aspect related to ambipolar diffusion: we consider whether there is 
mutual slippage between the magnetic field lines and the bulk plasma flow
in our experiments. To that end, the unsigned vertical 
flux, namely, 
\begin{equation}
    \Phi= \int^{x_f}_{x_0}{ |B_z|  dx}, 
\end{equation}
was computed for the four branches (LTE, LTE+AD, NEQ, NEQ+AD) of Experiment 1 and 2 at
different heights: $z=[0.5,1.0,2.0,3.0,4.0,5.0]$~Mm. The integration is calculated over the 
whole horizontal domain, so $x_0 = 0$~Mm and $x_f = 32$~Mm, for
15 minutes after the start of the branches at $t=60$ min. 
We also define $r_{\Phi}$ as the relative difference of the unsigned vertical flux between the branches with and without ambipolar diffusion:
    \begin{equation}
        r_{\Phi} = \frac{\Phi^{\rm AD}_{x} - \Phi_x}{\Phi_x}, \quad {\rm where}\  x = {\rm LTE, NEQ},
        \label{eq:ratio}
    \end{equation}
and compute the average, $< r_{\Phi}>$, and standard deviation, $\sigma_{r_{\Phi}}$, for 
those 15 minutes after the start of the branches.
The results are shown in Figure~\ref{fig:7} and Table~\ref{table:2}.

     Panel A of the figure shows the vertical
      magnetic flux for the LTE+AD branch, $\Phi^{\rm AD}_{\rm LTE}$, (colored
      solid lines) and the LTE case, $\Phi_{\rm LTE}$, (superimposed dashed
      lines) for different heights in Experiment 1. As can bee seen, $\Phi^{\rm AD}_{\rm LTE}$ and $\Phi_{\rm LTE}$ are almost identical,
      regardless of the height.  The relative difference between fluxes 
    is less than around 2\%. In
    addition, the standard deviation is a factor 2-3 of the mean value, 
    which indicates that there is a large variation in the ratio between both
    magnetic fluxes, with the LTE flux sometimes being 
    larger than the LTE+AD one, as seen for $z=0.5$ Mm, where $< r_{\Phi}>$ is negative.
    From this result, it seems that the ambipolar diffusion produces a negligible effect in the amount of emerged flux
    under the LTE assumption.
    
    The second panel from the top contains the magnetic flux for the NEQ+AD branch, $\Phi^{\rm AD}_{\rm NEQ}$, (colored solid lines) and the NEQ case, $\Phi_{\rm NEQ}$,  (superimposed dashed lines) for Experiment 1. We can again see that the magnetic flux in both cases is practically identical, no matter the height. In fact, the corresponding statistical values in Table~\ref{table:2} show that when considering NEQ, the relative 
 difference is even smaller than in LTE, finding that the average is close to
 0. The conclusion is that the magnetic slippage due to ambipolar
   diffusion, as measured by the $<r_\Phi>$ parameter, is also negligible when enabling the NEQ module. 
    Panel B of Figure~\ref{fig:7} also shows $\Phi^{\rm AD}_{\rm LTE}$ (gray curves) for comparison purposes between the LTE and NEQ
    cases. For this experiment, it seems that at the lower heights, $z=0.5$
    and $z=1.0$ Mm, the magnetic flux emerges sooner in NEQ than in LTE,
    while the opposite behavior is found for the other heights. In spite of
    this fact, at $t\approx 75$ min both NEQ and LTE fluxes converge to
    roughly the same value at the different heights, with the largest
    difference being found at $z=5.0$ Mm.
    
    The two remaining panels in  Figure~\ref{fig:7} (panels C and D) are the equivalent plots for Experiment 2. Those panels illustrate the same features explained for Experiment 1: the ambipolar diffusion has no significant impact on the emerged magnetic flux, neither in NEQ nor in LTE, which can be also checked through the corresponding statistical values in Table~\ref{table:2}.

\begin{table*}[!ht]
\caption{Statistical values for the average $< r_{\Phi}>$ (see Equation \ref{eq:ratio}) and corresponding standard deviation, $\sigma_{r_{\Phi}}$ , for flux curves plotted at different heights $z$ in Figure \ref{fig:7}. The
values have been multiplied by $10^2$ to have them in percentages.}
\label{table:2}      
\centering          
\begin{tabular}{|| c | c | c | c || c | c | c | c ||}     % 8 columns 
 \hline
 \multicolumn{4}{||c||}{Experiment 1} & \multicolumn{4}{|c||}{Experiment 2} \\
 \hline
\hline\hline
\rule{0pt}{2.1ex} 
Branches & $z$ (Mm) & $<r_{\Phi}>$ (\%) & $\sigma_{r_{\Phi}}$ (\%)  & Branches &  $z$ (Mm) & $<r_{\Phi}>$ (\%) & $\sigma_{r_{\Phi}}$ (\%)   \\

 %%%%%%%%%%%%%%%%%%%%%%%%%%%%%%%%%%%%%%%%%%%%%%%%%%%%%%%%%%%%%%%%%%%%%%%%%%%%%%%%%%%%%%%%%%%%%%%%%%%%%%%%%%%%%%%%%%%%%%%%%%%%%%
 % LTE
 %%%%%%%%%%%%%%%%%%%%%%%%%%%%%%%%%%%%%%%%%%%%%%%%%%%%%%%%%%%%%%%%%%%%%%%%%%%%%%%%%%%%%%%%%%%%%%%%%%%%%%%%%%%%%%%%%%%%%%%%%%%%%%
\hline\hline
\rule{0pt}{2.0ex} 
\multirow{6}{*}{$\frac{\rm LTE+AD}{\rm LTE}$
} & 0.5 & -0.2 & 0.5 & \multirow{6}{*}{$\frac{\rm LTE+AD}{\rm LTE}$}  & 0.5 & 0.4  &   0.7 \\ 
%\hline 
 & 1.0 & 0.2   &  0.7   & & 1.0 & 0.0   &  0.9 \\
%\hline     
  & 2.0 & 1.7   &   2.7  &  & 2.0 & 2.7 &  4.3    \\
%\hline     
  & 3.0 &  2.4   &   6.1  &  & 3.0 & 2.5  &   4.2   \\
%\hline     
  & 4.0 & 2.0  &   5.6&    & 4.0 & 2.8   &   5.7  \\
%\hline     
  & 5.0 & 1.8   &  5.4 &  & 5.0 & 3.6  &   7.9   \\
 %%%%%%%%%%%%%%%%%%%%%%%%%%%%%%%%%%%%%%%%%%%%%%%%%%%%%%%%%%%%%%%%%%%%%%%%%%%%%%%%%%%%%%%%%%%%%%%%%%%%%%%%%%%%%%%%%%%%%%%%%%%%%%%
 % NEQ
 %%%%%%%%%%%%%%%%%%%%%%%%%%%%%%%%%%%%%%%%%%%%%%%%%%%%%%%%%%%%%%%%%%%%%%%%%%%%%%%%%%%%%%%%%%%%%%%%%%%%%%%%%%%%%%%%%%%%%%%%%%%%%%%
\hline\hline
\rule{0pt}{2.0ex} 
\multirow{6}{*}{$\frac{\rm NEQ+AD}{\rm NEQ}$}
  & 0.5 & -0.2 & 0.5 & \multirow{6}{*}{$\frac{\rm NEQ+AD}{\rm NEQ}$} & 0.5 & -0.2 &    1.7 \\ 
%\hline 
  & 1.0 & -0.2 &  0.6 &   & 1.0 & -0.1  &   0.8 \\ 
%\hline     
  & 2.0 & -0.6  &   1.5 &   & 2.0 & -0.7  &   1.2   \\ 
%\hline     
  & 3.0 & -0.2   &   1.3 &   & 3.0 & -1.4   &  2.0 \\ 
%\hline     
  & 4.0 & -0.3   &   1.2 &   & 4.0 & -1.4  &  2.6\\ 
%\hline     
  & 5.0 & -0.2  &   1.1  &   & 5.0 & -2.0  &   4.0 \\ 
\hline  
\end{tabular}
\end{table*}

A test has also been carried out checking for variations in the amount
  of chromospheric plasma that rises during the magnetic flux emergence
  process. To that end, we have performed the same kind of analysis
  like  for the magnetic flux: we have calculated the integral
\begin{equation}
    \Psi = \int^{x_f}_{x_0}{ \rho \, dx},
\end{equation}
for the same heights as before, namely
$z=[0.5,1.0,2.0,3.0,4.0,5.0]$~Mm. The quantity $\Psi$ is the
  integrated column density along the given horizontal level; monitoring
  $\Psi$ may help discover any deviation in the amount of emerged plasma
  between the different cases. In fact, we find that there is no significant
difference in the amount of mass lifted by the emergence process when
including the ambipolar diffusion or otherwise, which is a natural
consequence of the lack of magnetic field slippage previously found.

In order to understand the aforementioned 
lack of magnetic field slippage, we
have to take a look at the drift velocity $\uamb$, which
measures the departure of the ion velocity from the 
bulk plasma velocity $\mathbf{u}$ and is given by
\begin{equation}
    \uamb = \etaamb \frac{({\mathbf J} \times {\mathbf B})}{|\mathbf{B}|^2} = \etaamb \frac{( \hat{\mathbf{j}} \times \hat{\mathbf{b}})}{ L_B} =  
    \frac{|\bf{B}|^2}{\rho^2} \frac{\rho_N^2}{\Sigma_n \Sigma_i \rho_n \nu^{\ast}_{ni}} \frac{( \hat{\mathbf{j}} \times \hat{\mathbf{b}})}{ L_B}.
    \label{eq:u_amb}
\end{equation}
The order of magnitude of $|\uamb|$ is thus given by three factors:
  (a) the square of $|\mathbf{B}|/\rho$; (b) a somewhat complicated
  expression whose spatial variation, however, in most cases is just given by
  that of the ratio   $\rho_n/\rho_i$; and (c) the inverse of the characteristic
  length of the 
  magnetic field $L_B^{-1} \equiv \left|(\nabla \times B)/B\right|$. 
The module of the remaining term, $|\hat{\mathbf j} \times \hat{\mathbf b}|$,
is expected to be of order one, since the electric
current in the emerged domes is nearly perpendicular to the magnetic
field. In the following, we analyze these blocks using Experiment 2 to
illustrate with examples.

\begin{figure*}[!ht]
    \centering
    \includegraphics[width=1.00\textwidth]{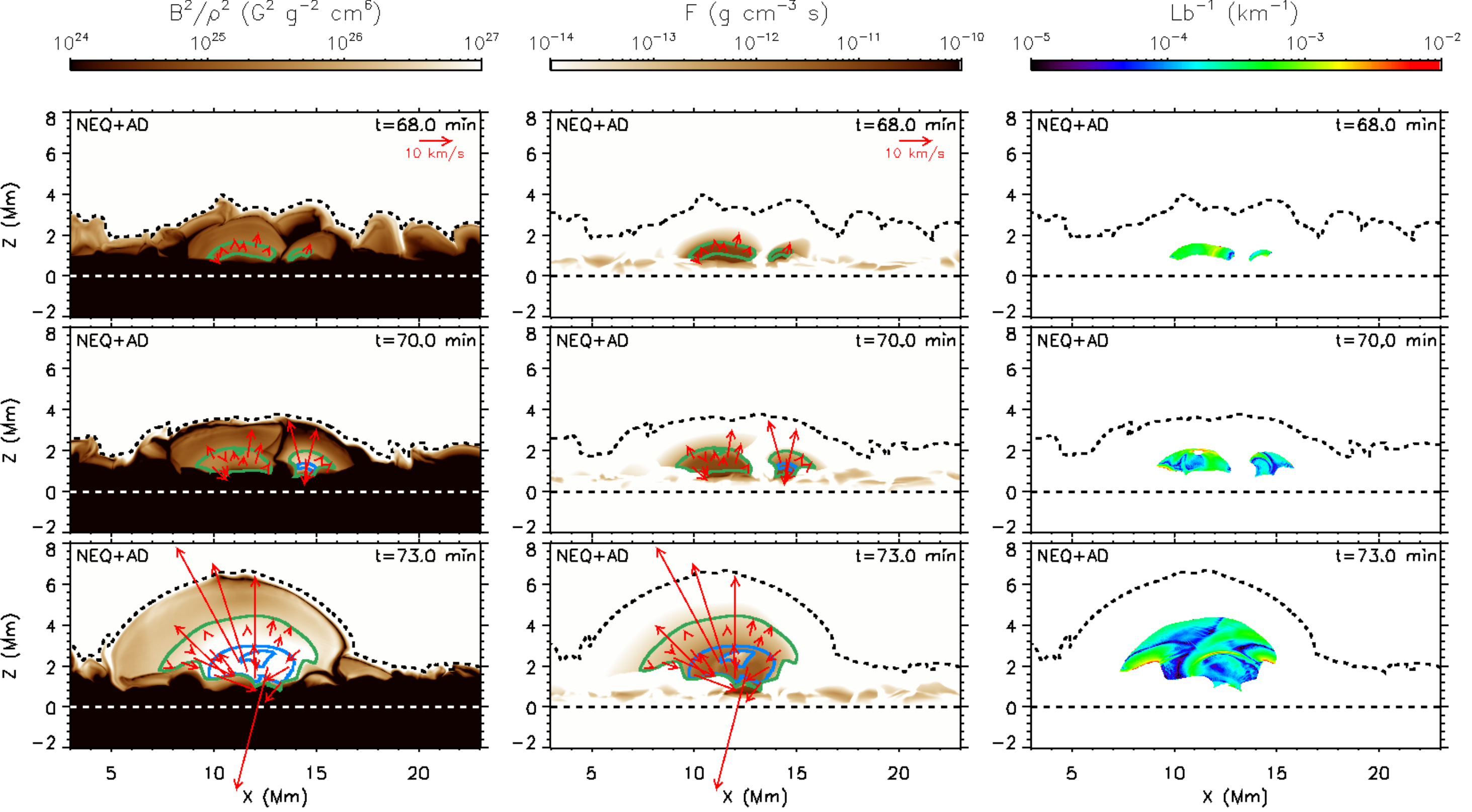}
    \caption{The various components that determine the drift 
velocity $\uamb$, calculated for Experiment 2. 
    Left column: Ratio $|\bf{B}|^2/\rho^2$. 
    Middle column: maps of $F=\rho_N^2/(\Sigma_n \Sigma_i \rho_n \nu^{\ast}_{ni})$.
    Right column: Inverse of the characteristic length of the magnetic field, $L_B^{-1} \equiv \left|(\nabla \times B)/B\right|$ only in places where $\etaamb > 10^{13}$ cm$^2$ s$^{-1}$.
    Two isocontours for $\etaamb$ are superimposed in the left and middle column panels as green lines ($10^{13}$ cm$^2$ s$^{-1}$) 
    and blue lines ($10^{15}$ cm$^2$ s$^{-1}$). The velocity field due to the ambipolar diffusion $\uamb$ 
    (see Equation \ref{eq:u_amb}) is also shown in those two columns with red arrows only in regions
    where $\etaamb > 10^{13}$ cm$^2$ s$^{-1}$. Additionally, dashed lines
    indicate the location of  the solar surface (line at
    $z=0$ Mm) and of the transition region (isocontour at $T=10^5$ K).
    An animation of this figure is available online.}
    \label{fig:8}  
\end{figure*}

    The left column of Figure \ref{fig:8} shows the ratio
      $|\bf{B}|^2/\rho^2$ at three different stages of the flux emergence
      process. At earlier stages, we see a moderate spatial variation of this factor within the emerged dome by, at most, about an order of
      magnitude. Later, larger values of this ratio are found mainly due to the decrease in density as the dome expands and gets rarefied.

    The middle term of Equation \ref{eq:u_amb}, $\rho_N^2/(\Sigma_n
      \Sigma_i \rho_n \nu^{\ast}_{ni})$, is shown in the central column of
      Figure~\ref{fig:8}. In the figure, one can see that the term is larger
      in the core of the dome than in the periphery during almost the whole
      evolution of the emerged dome, in fact by at least a few orders of
      magnitude. The reason for this is that the core of the dome is much
      less ionized than its periphery and, as already said, that term follows
      approximately the ratio $\rho_n/\rho_i$.

   The two factors studied so far, when combined, give the
      value of the ambipolar diffusion coefficient. From the results just
      explained, one would expect $\etaamb$ to be larger toward the core of
      the dome and decreasing toward the periphery at the top and
      sides. This is indeed the case, as shown through the solid solid
      isocontours in the left and middle columns of Figure~\ref{fig:8}, which
      correspond to $\etaamb=10^{13}$ cm$^2$ s$^{-1}$ (green) and $\eta_{\rm
        amb}=10^{15}$ cm$^2$ s$^{-1}$ (blue).  Through those isocontours,
      we can also see that $\etaamb$ increases as the dome
        develops, specially in the innermost part. Therefore, during the
      first stages of the emergence, obtaining magnetic field slippage is
      less probable; in later stages, if any slippage occurs, it is likely to
      be located in the core of the emerged region.

    Concerning the characteristic length of the magnetic field, our
      emerged regions are not simple idealized symmetric magnetic domes.
      This leads to a large complexity in the magnetic field structure that
      emerges. In order to show this, in the right column of Figure
      \ref{fig:8}, we have plotted $L_B^{-1}$ only in regions where the
      ambipolar diffusion coefficient is $\etaamb>10^{13}$~cm$^2$~s$^{-1}$.
      In those panels, it is possible to find regions with variations in
      $L_B^{-1}$ up to a few orders of magnitude, consequently meaning large
      variations in the module of the drift velocity $\uamb$.

Additionally to the module, we can analyze the direction of
  $\uamb$, given by the vector product $( \hat{\mathbf{j}} \times
  \hat{\mathbf{b}})$. For compactness, in the left and middle columns of
  Figure \ref{fig:8} we have superimposed a set of red arrows
        to show the $\uamb$ vector field in regions
      where $\etaamb>10^{13}$ cm$^2$ s$^{-1}$. The arrows do not show a clear
      pattern; sometimes even showing opposite directions in the
      velocity drift. In addition, in the accompanying movie, we find rapid variations of those
      directions, which reflects the highly dynamic environment of the
      emerged regions.

\begin{figure}[!ht]
    \centering
    \includegraphics[width=0.46\textwidth]{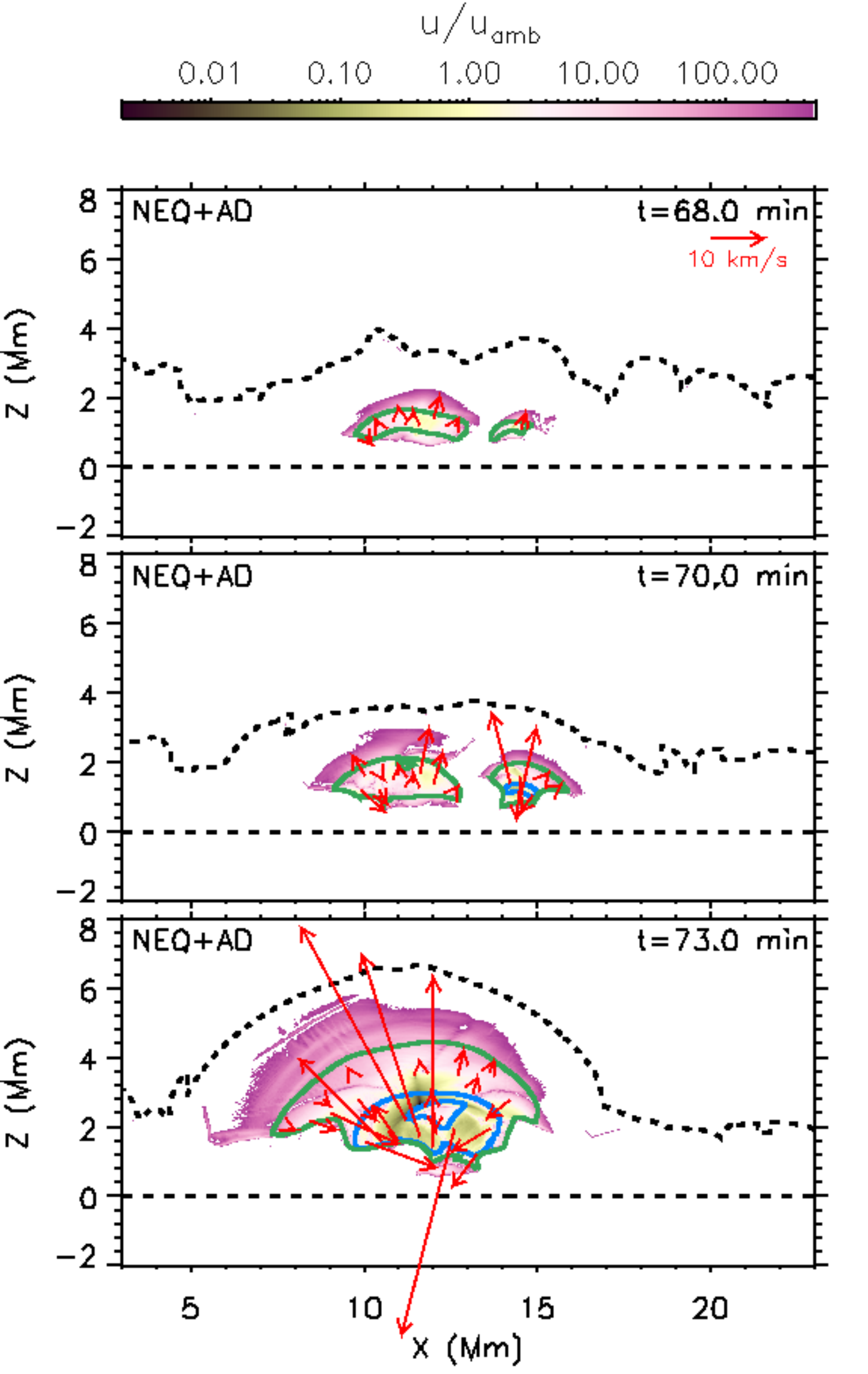}
    \caption{Maps of $|\mathbf{u}| / |\uamb|$ in the plane XZ. 
    Two isocontours for $\etaamb$ are superimposed in the left and right column panels as green lines ($10^{13}$ cm$^2$ s$^{-1}$) 
    and blue lines ($10^{15}$ cm$^2$ s$^{-1}$). The velocity field due to the ambipolar diffusion $\uamb$ 
    (see Equation \ref{eq:u_amb}) is also shown in those columns with red arrows only in regions
    where $\etaamb > 10^{13}$ cm$^2$ s$^{-1}$. Additionally, dashed lines
    indicate the location of  the solar surface (line at
    $z=0$ Mm) and of the transition region (isocontour at $T=10^5$ K).
    An animation of this figure is available online.}
    \label{fig:9}  
\end{figure}

 Once we know the dependencies of the module and direction of
  $\uamb$, the next step is to analyze how large 
  this drift velocity is in comparison with the plasma velocity
  $\mathbf{u}$.  To that end, the ratio $|\mathbf{u}| / |\mathbf{u}_{\rm
    amb}|$ contained in the XZ-plane is studied. Figure \ref{fig:9} contains the
  results for the NEQ+AD branch of Experiment 2. The first thing we notice is that this ratio is
  only equal or smaller than $1$ in some particular regions and that their
  corresponding area is smaller than the surface of the magnetized dome. In the accompanying animation of Figure
  \ref{fig:9}, we find that the regions where $|\uamb|$ dominates
change faster than the characteristic times of the dome expansion;
those regions also show rapid variations of the direction of
  $\uamb$ (red arrows). This means that the regions
  where field slippage could appear are small, and show rapid variations of
  the drift velocity in module and direction. As a consequence, no
significant slippage (and therefore plasma material leakage) is found in our
numerical experiments.

%____________________________________________________________________________________________________________________________________
%
% SECTION 5:  Conclusions and discussion
%
%____________________________________________________________________________________________________________________________________
\section{Conclusions and discussion}\label{sec:conclusions}

We have performed two 2.5D radiative-MHD numerical experiments of magnetic flux emergence
from the upper layers of the convection zone to the corona. The experiments were  
carried out using the Bifrost code, including an extra module that computes the 
NEQ formation of atomic and molecular hydrogen and another one that considers the
partial ionization effects, in particular, the ambipolar diffusion. The time
evolution of all the experiments leads to the formation of emerged magnetized 
regions with shapes similar to domes. The relevance of
the NEQ and ambipolar diffusion effects were studied within those emerged domes
as well as the role of H$_2$ molecules and shocks in the thermodynamics
of the magnetized domes.  Furthermore, we have compared the amount of unsigned 
magnetic flux that emerges when including the NEQ ionization and recombination and ambipolar 
diffusion effects or otherwise. In the following, we summarize the main conclusions about the effects of the NEQ  ionization and recombination
and ambipolar diffusion when dealing with magnetic flux emergence, as well as some possible implications for other
  chromospheric environments.

    %LTE vs NEQ for the ambipolar diffusion
    In Section \ref{sec:emerged_region_properties}, it was shown that the LTE assumption can lead to an important underestimation of the ionization fraction in the emerged region, specially in the upper part where the difference can be up to 2-3 orders of magnitude. This has direct consequences on the ambipolar diffusion term since it inversely depends on the number of ions: assuming LTE can highly overestimate the effects of the ambipolar diffusion. The consequences of the NEQ ionization and recombination of hydrogen for the ambipolar diffusion may also have an impact on other chromospheric contexts, so one should be careful about the results that have been obtained under the LTE assumption.

    %The importance of H2  
    This work shows that the formation rate of the H$_2$ molecule
    is key in the thermodynamics of the emerged region. Under the LTE
    assumption, we find around 60-80\% of H$_2$ molecules
    by number 
    within the emerged domes, while for the NEQ cases the maximum value is
    around 27\%. The right determination of the number of H$_2$ molecules is
    important to properly calculate their contribution to the energy due to
    their exothermic formation. Through observations, estimations of the
    fraction of H$_2$ have been inferred in various sunspot umbrae using the
    equivalent width of the OH 15652 \AA\ line as a proxy
    \citep{Jaeggli:2012}; they find that substantial H$_2$ molecule formation
    is present. However, the question remains open whether
      it would be possible to get
    similar estimates from observations of magnetic flux emergence regions.

    %The minimum temperature 
    The NEQ+AD branch of our two experiments shows that the strong cooling
    due to the expansion of the magnetized plasma during the emerging process
    can not be totally counteracted by the ambipolar diffusion
    heating. Consequently, artificial heating is still
    necessary to prevent even cooler temperatures than our threshold at
    $1660$~K, 
    where the radiation and EOS tables  of  the  code  are  not  accurate.
    The minimum temperature reached in flux emergence processes thus remains an open question and adds to the question of the minimum temperature in the quiet sun discussed by \cite{Leenaarts:2011}.

    %Shocks and ambipolar diffusion heating
    In Section \ref{sec:tauqjamb}, we could see that the emerged region is continually traversed by multiple shocks that have a significant impact on the thermodynamics of the magnetized domes. In particular, when considering the ambipolar diffusion in the NEQ branch, the temperature is up to a factor two greater than without ambipolar diffusion, while the ionization fraction can be up to one order of magnitude greater. The reason is the efficiency of the heating by ambipolar diffusion in the shocks, with characteristic times that range from $0.1$ to $100$~s. This heating mechanism can help to offset some of the strong cooling during the expansion of the emerging magnetized plasma. In this vein, by means of non-LTE inversions of observations, \cite{Leenaarts:2018} have found evidences of heating in the chromosphere associated with flux emergence. The authors conjecture two possible sources for this heating: one associated with current sheets and a more homogeneous one due to ambipolar diffusion. However, those authors do not mention shocks as a possible source and we have not found homogeneous heating by ion-neutral collisions.

\begin{figure}[!ht]
    \centering
    \includegraphics[width=0.45\textwidth]{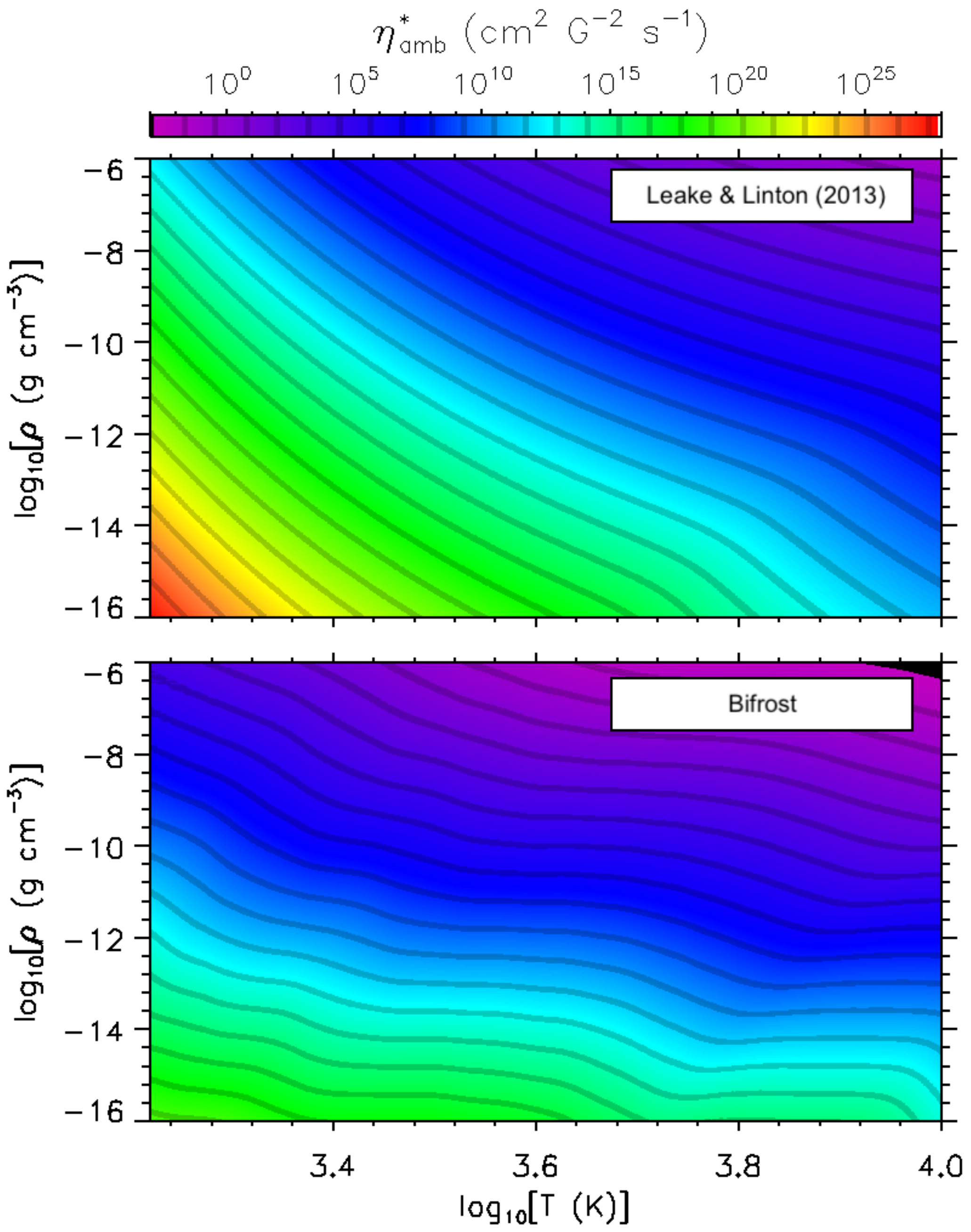}
    \caption{Map of $\etaambb$ (see Equation \ref{eq:eta_ambb}) as a function of density and temperature. 
    Top panel contains $\etaambb$ calculated using the same EOS than \cite{Leake:2013dq}. Bottom panel illustrates $\etaambb$
    computed with the Bifrost EOS that we use for the LTE branches.}
    \label{fig:10}  
\end{figure}

    %Magnetic field slippage 
    In Section \ref{sec:slippage}, we analyzed in
    detail the effects of the ambipolar diffusion on the amount of emerged
    magnetic flux and lifted mass. We found that during the first stages after the buoyancy
    instability, mutual slippage of plasma and magnetic field is unlikely to
    take place due to the small values of the ambipolar diffusion
    coefficient, and therefore of the velocity drift of the ions. Later, the
    ambipolar diffusion increases, specially in the innermost part of the
    dome, and so does the velocity drift, as a consequence of the creation of
    more neutrals during the expansion of the magnetized
    plasma. However, the direction of the drift speed changes
      rapidly. On top of that, the regions where the drift can be important
      are a small fraction of the total area covered by the new magnetic flux
      and they are non-stationary. Consequently, the ambipolar diffusion does not
      significantly change the amount of emerged magnetic flux (as seen in 
    Figure \ref{fig:7} and Table \ref{table:2}), nor, consequently 
    the amount of lifted mass, irrespective of the assumption of equilibrium equilibrium (NEQ or LTE).
      We have compared this result with the related ones in the literature. For instance,  
      \cite{Leake:2006kx} found that the rate of magnetic flux emergence is considerably increased
        when including ambipolar diffusion in the calculation. \cite{Arber:2007yf} found that the main difference
        between fully ionized 3D calculations and partially ionized ones is that
        the former are able to lift more chromospheric plasma.
        More recently, \cite{Leake:2013dq} improved the EOS used in the two previous papers
        by including changes in the internal energy density due to
        ionization and recombination. They showed that this change in the EOS significantly decreases the amount of \textit{out-of-plane} magnetic flux (or shear flux) and
        mass lifted during the flux emergence process as compared to results with the previous EOS; however, they still find that simulations 
        without ambipolar diffusion raise between 7.7 and 10 times more mass.
        We think that the main reason for the discrepancy with our results comes from the difference between the EOS. 
    Those authors consider a plasma consisting of pure hydrogen. This means that as
    the temperature drops, the number of ions tends to zero due to the lack
    of other ionized elements that also contribute with electrons, resulting
    in large values of the ambipolar diffusion because $\etaamb \propto
    n_i^{-1}$. In addition, those authors do not include the H$_2$ molecule
    formation, so they do not have the pool of energy related to its
    exothermic formation that can also affect the thermodynamics. In order to
    show how important the choice of EOS is, Figure \ref{fig:10} shows the
    ambipolar diffusion coefficient normalized to the magnetic field:
   \begin{equation}
        \etaambb = \frac{\etaamb}{|\mathbf{B}|^2}.
        \label{eq:eta_ambb}
  \end{equation}
  In the top panel of the figure, $\etaambb$ is calculated as a
    function of density and temperature following the EOS
  used by \cite{Leake:2013dq}; in the bottom panel, following the Bifrost EOS
  used for the LTE branches 
  of this paper. The figure clearly shows that
  considering only hydrogen can lead to a substantial overestimate
  of the role of the ambipolar diffusion. 

Through the results of this paper, we conclude that the NEQ imbalance 
in the populations of hydrogen together with ambipolar diffusion effects 
have important consequences for the dynamics and thermodynamics 
of magnetic flux emergence processes. 
Future research must explore the consequences of NEQ ionization and recombination 
in the ion populations of other elements. 
Also, further effects, like the Hall term in the Generalized Ohm's Law, must be studied.

\appendix
\section{Choosing the instant to create the branches}\label{sec:appendix}
One of the concerns when carrying out the present work was choosing the right instant to create the different branches. In this Appendix, we prove that the NEQ and AD effects are negligible within the emerging magnetized plasma in the instants prior to our branching at $t=60$ min, meaning
 that the results presented in this paper are not affected by our choice.

\begin{figure}[!ht]
    \centering
    \includegraphics[width=0.45\textwidth]{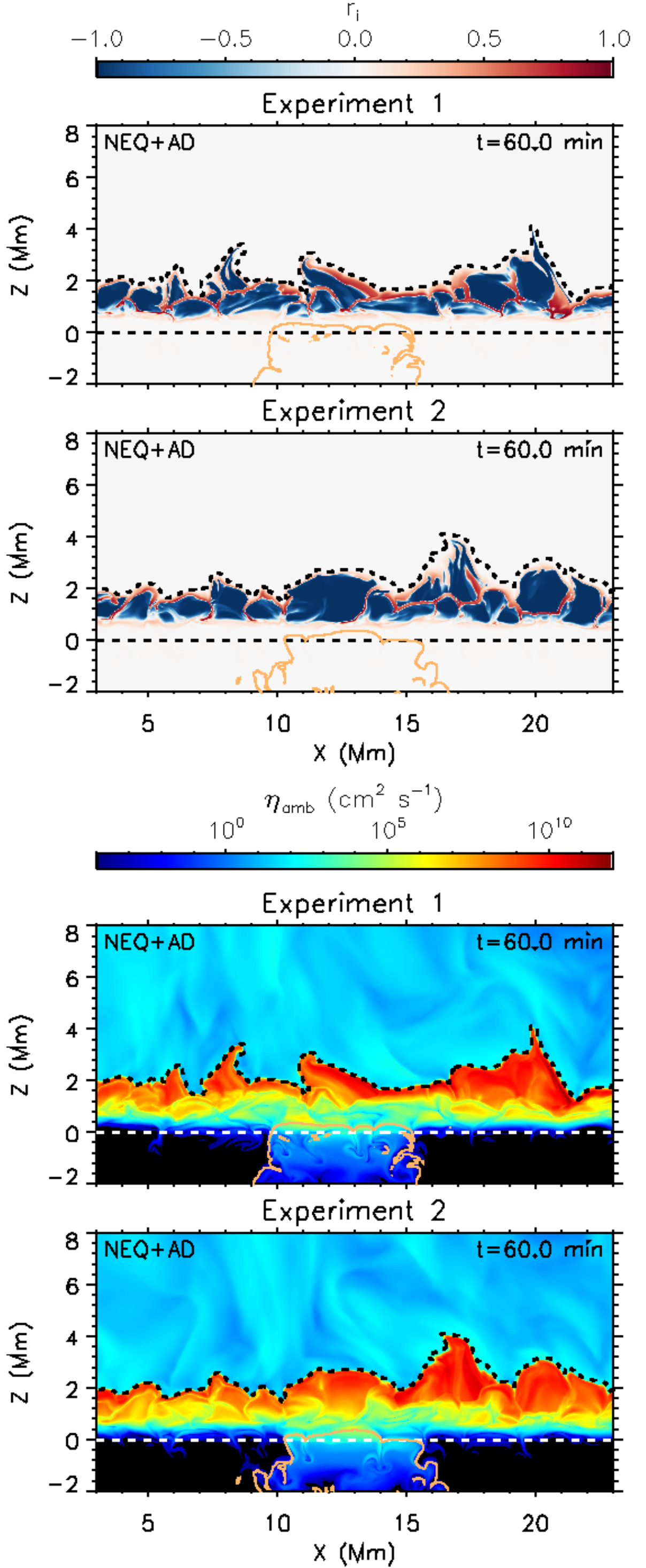}
    \caption{Maps of $r_i$ (see Equation \ref{eq:r_i}) and $\etaamb$ for Experiments 1 and 2 at $t=60$ min. 
    The location of the new emerging magnetic field is shown through a solid isocontour at $|\bf{B}|$ $=10$ G in orange color. 
    Additionally, dashed lines indicate the location of  the solar surface (line at
    $z=0$ Mm) and of the transition region (isocontour at $T=10^5$ K).
    An animation of this figure showing the evolution from $t=50$ to $t=60$ min is available online.}
    \label{fig:a1}  
\end{figure}

    %NEQ
    Concerning the NEQ effects, we compare the ion density in the NEQ+AD branch, $\rho_i$,  with
    the ion density obtained from an LTE calculation based on the values of that experiment, $\rho^{*}_i$, 
    using the following relative difference:
    \begin{equation}
       r_i = \frac{\rho^{*}_i - \rho_i}{\rho^{*}_i + \rho_i}.
    \label{eq:r_i}
    \end{equation}  
    This ratio is shown for Experiments 1 and 2 at $t=60$ min in the two uppermost panels of Figure \ref{fig:a1}, and from $t=50$ to $t=60$ min in
    the associated movie.
    In both panels, the location of the new emerging magnetic field is shown through a solid isocontour at $|\bf{B}|=10$ G in orange color. 
    The values of $r_i$ clearly indicate that the departures in the ion density from ionization equilibrium occur higher up in the atmosphere and that $t=60$ min is a safe point in which we can turn off the NEQ module without having already affected the emergence of the tube.
    
    %AD
    Regarding the AD effects, 
    we have plotted the ambipolar diffusion coefficient for Experiments 1 and 2 at $t=60$ min in the two lowermost panels of Figure \ref{fig:a1}. As in the previous panels, the location of the new emerging magnetic field is shown by means of a solid isocontour at $|\bf{B}|$ $=10$ G in orange color. 
    During most of the emergence through the convection zone (see associated animation), the coefficient $\etaamb$ in the twisted tube is $\leq10^2$ cm$^2$ s$^{-1}$, which is 
    smaller or comparable to the negligible values found in the corona. As the magnetized plasma reaches the surface and piles up, there is an increase in the 
    ambipolar diffusion coefficient up to values around $10^5-10^6$ cm$^2$ s$^{-1}$. These values are still 8-9 orders of magnitude smaller 
    than the values we find later within the emerged region once it reaches the chromosphere and expands.  
    As a consequence, we can create the branches without AD also at $t=60$ min without having already affected the emergence of the tube.

%__________________________________________________________________
%__________________________________________________________________
\begin{acknowledgements}
This research is supported by the Research Council of Norway through its
Centres of Excellence scheme, project number 262622, and through grants of
computing time from the Programme for Supercomputing. It is also supported by 
the Spanish Ministry of Science, Innovation and Universities through
projects AYA2014-55078-P and PGC2018-095832-B-I00, as well as through the
Synergy Grant number 810218 (ERC-2018-SyG) of the European Research Council; 
by NASA through 
grants NNX16AG90G, NNX17AD33G, 80NSSC18K1285 and contract NNG09FA40C (\IRIS) and by the NSF grant AST1714955.
The authors thankfully acknowledge the computer
resources provided at the Pleiades cluster through the computing projects s1061,
s1630, and s2053 from the High End Computing (HEC) division of NASA.
In addition, this study has been discussed within the activities of team 399
``Studying magnetic-field-regulated heating in the solar chromosphere’’ at the
International Space Science Institute (ISSI) in Switzerland.

\end{acknowledgements}

%-------------------------------------------------------------------

\bibliographystyle{aa}
\bibliography{collectionbib}

\end{document}